\documentclass[twocolumn,conference,letterpaper]{IEEEtran}
\usepackage[left=0.75in,right=0.75in,top=1in,bottom=0.75in]{geometry}

\usepackage{cite}
\usepackage{amssymb,amsmath,latexsym,amsfonts,amsthm,mathtools}

\usepackage{graphicx}
\usepackage{float,subcaption}
\usepackage{textcomp}
\usepackage{xcolor}
\definecolor{darkpastelpurple}{rgb}{0.59, 0.44, 0.84}
\usepackage{hyperref}

\usepackage{algorithm}
\usepackage{algpseudocode}

\hypersetup{colorlinks, breaklinks, citecolor=blue, linkcolor=blue, urlcolor=blue}
\usepackage[nameinlink]{cleveref}
\Crefformat{figure}{#2Fig.~#1#3}
\Crefmultiformat{figure}{Figs.~#2#1#3}{ and~#2#1#3}{, #2#1#3}{ and~#2#1#3}

\usepackage{tikz}
\usetikzlibrary{automata, shapes, arrows, calc, arrows.meta, fit, positioning}

\theoremstyle{plain}
\newtheorem{theorem}{Theorem}
\newtheorem{lemma}{Lemma}

\newtheorem*{problem*}{Problem}
\theoremstyle{remark}
\newtheorem{remark}{Remark}

\theoremstyle{definition}

\newcommand{\sign}{\mathrm{sign}}
\usepackage{mathrsfs}

\usepackage{newtxmath}
\usepackage{booktabs}
\usepackage{duckuments}

\IEEEoverridecommandlockouts

\begin{document}
	\title{Self-organizing Multiagent Target Enclosing under Limited Information and Safety Guarantees}
	\author{Praveen Kumar Ranjan, Abhinav Sinha,~\IEEEmembership{Member,~IEEE}, Yongcan Cao,~\IEEEmembership{Senior Member,~IEEE}
\thanks{P. K. Ranjan and Y. Cao are with the Unmanned Systems Lab, Department of Electrical and Computer Engineering, The University of Texas at San Antonio, TX- 78249 USA. Emails: praveen.ranjan@my.utsa.edu, yongcan.cao@utsa.edu.\newline A. Sinha is with the GALACxIS Lab, Department of Aerospace Engineering and Engineering Mechanics, University of Cincinnati, OH- 45221 USA. Email: abhinav.sinha@uc.edu.}
}

	\maketitle
	\thispagestyle{empty}
	
	\begin{abstract}
    This paper introduces an approach to address the target enclosing problem using non-holonomic multiagent systems, where agents self-organize on the enclosing shape around a fixed target. In our approach, agents independently move toward the desired enclosing geometry when apart and activate the collision avoidance mechanism when a collision is imminent, thereby guaranteeing inter-agent safety. Our approach combines global enclosing behavior and local collision avoidance mechanisms by devising a special potential function and sliding manifold. We rigorously show that an agent does not need to ensure safety with every other agent and put forth a concept of the nearest colliding agent (for any arbitrary agent) with whom ensuring safety is sufficient to avoid collisions in the entire swarm. The proposed control eliminates the need for a fixed or pre-established agent arrangement around the target and requires only relative information between an agent and the target. This makes our design particularly appealing for scenarios with limited global information, hence significantly reducing communication requirements. We finally present simulation results to vindicate the efficacy of the proposed method.
	\end{abstract}
	

\section{Introduction}    
    Autonomous vehicles have gained significant prominence in a range of tasks, including reconnaissance, surveillance, crop and forest monitoring, and search and rescue missions, e.g., \cite{doi:10.2514/1.G007057,5417166, 9329129,8306922,10463136}. Most of these tasks share a common behavior where an autonomous vehicle, known as the pursuer, monitors an object of interest, known as the target. In general, the pursuer's stable motion around the target by maintaining the desired separation from the target is referred to as target enclosing \cite{10251969} or encirclement/circumnavigation/entrapment when the enclosing orbit is a circle \cite{9924233}. 
    
    One of the earliest approaches for target circumnavigation entails pursuers confining the target in a specific geometric shape posing as a formation control problem. The authors in \cite{KIM20071426, MARSHALL20063} utilize a cyclic pursuit strategy to cooperatively enclose a target, achieving and maintaining uniform angular spacing among the agents. Vector field guidance is another popular approach, where vector fields can be designed to create a limit cycle or periodic orbit around the target, guiding agents to desired geometric shapes \cite{doi:10.2514/1.30507,doi:10.2514/1.G002281}. However, each orbit may require a separate field, thus limiting its applicability in a multiagent scenario. Several other guidance strategies have also been developed for trapping targets in circular orbits, based on differential geometry \cite{rendev2013}, model predictive control \cite{7066874}, and reinforcement learning \cite{SHAO2023108609}. Many of the works in the target enclosing have focused on reducing the information required for developing the control laws. For example, some strategies utilize only range-only measurements \cite{CAO2015150, DONG2020108932,9613761}, bearing angle measurement \cite{doi:10.2514/1.G001421,doi:10.2514/1.G002707,iet-cta.2018.6133}, range and bearing information \cite{doi:10.2514/1.G006403,9924233,doi:10.2514/6.2024-0124}. In \cite{doi:10.2514/1.G007539,doi:10.2514/1.G006957}, authors developed a flexible guidance strategy that significantly enhances the pursuer's maneuverability by allowing it to enclose the target in arbitrary shapes.
    
    While most of the aforementioned studies focus on developing guidance strategies for a single pursuer, many researchers have explored cooperative targets enclosing \cite{KIM20071426,MARSHALL20063,9705074,SHI2021104873,9440725}. These approaches aim to reduce the complexity of guidance strategies by requiring every pursuer to coordinate with only some of the neighboring agents. The authors in \cite{KIM20071426,MARSHALL20063} designed a guidance strategy that arranges the pursuers in a circular formation, achieving uniform angular spacing on the orbit around the target. The works in \cite{9705074,SHI2021104873,9440725}, utilized various rigid formations between the pursuers and the target converging to desired relative configurations on the enclosing shape. Note that most of the above-mentioned works did not account for explicit inter-agent safety and focused on guiding the pursuers to the enclosing geometry without considering the possibility of collisions. Although a rigid formation shape may inherently ensure that no two pursuers occupy the same position at the steady state, additional considerations are necessary to prevent collisions at all times.
    
    Therefore, recent works have incorporated collision-avoidance guarantees in target-enclosing strategies \cite{LAN2010381,franchi2016decentralized,10234074,9424952,8771188,10195223}, steering the pursuers on a collision-free trajectory towards the enclosing shape around the target. For example, in \cite{LAN2010381,franchi2016decentralized,9424952}, pursuers achieve uniform angular spacing, while \cite{10234074,8771188} allow for predefined uneven angular spacing among pursuers on a fixed orbit around the target. The guidance strategy in \cite{10195223} not only avoids collisions among pursuers but also with static obstacles. Note that all the above works on multi-agent target enclosing utilize a rigid formation structure that significantly limits each pursuer's maneuverability and necessitates vehicle-to-vehicle communication. Each pursuer must simultaneously meet formation objectives (desired position and orientation) relative to both the target and multiple other pursuers to maintain the desired formation shape. In practice, pursuers may not be able to accurately track the desired position due to external disturbances or steady-state errors. Therefore, a strict formation structure may not be suitable for multi-agent target enclosing, as it constrains the pursuers to maneuver in accordance with multiple pursuers, and the failure of even a single pursuer may be catastrophic for the entire formation.
    
    This work is motivated by the need to develop safe motion coordination strategies for target enclosing that do not rely on rigid formation structures. Traditional target enclosing strategies can be overly restrictive, limiting the maneuverability of individual pursuers and requiring constant communication to maintain precise positions. This rigidity not only increases the computational and communication burden on the system but also reduces its robustness and adaptability in dynamic environments. The proposed flexible coordination strategy offers significant advantages as it allows pursuers to maneuver more freely, adapting to changing conditions and unexpected obstacles (in the form of neighboring pursuers) without compromising the overall goal of the target enclosing. By reducing the adherence to strict formation maintenance, the proposed strategy also minimizes communication efforts and the need for extensive prior information about the target and the environment. This can lead to more efficient and scalable solutions, particularly in scenarios where the environment is contested or the communication infrastructure is limited. The contributions of this paper can be summarized as follows:   
    \begin{enumerate}
        \item We develop a distributed guidance law for the pursuers to safely enclose a stationary target,  allowing them to autonomously arrange themselves on the enclosing orbit. This work builds on our previous research \cite{10251969,9924233,CAO2015150,doi:10.2514/6.2024-0124,doi:10.2514/1.G007539,doi:10.2514/1.G006957}, where we developed flexible enclosing strategies for a single pursuer-target pair. We extend those strategies to accommodate multiple pursuers, ensuring explicit safety guarantees. Unlike previous methods on multi-agent target enclosing \cite{KIM20071426,MARSHALL20063,9705074,SHI2021104873,9440725,LAN2010381,franchi2016decentralized,10234074,9424952,8771188,10195223} that rely on fixed or predetermined formations that restricts the final configuration of pursuers about the target, our approach enables self-organizing behavior. Such a behavior inherently allows the enclosing formation to adapt to a changing number of pursuers, with vehicles ensuring a safe distance among them. 
        \item Our approach steers the pursuers to arbitrary points on the enclosing orbit and ensures they do not collide with each other. This allows them to achieve cooperative target enclosing with any final relative configuration and eliminates the need for multiple pursuers to coordinate maneuvers to avoid collisions. Consequently, our method significantly enhances the maneuverability of each pursuer compared to previous works \cite{KIM20071426,MARSHALL20063,9705074,SHI2021104873,9440725,LAN2010381,franchi2016decentralized,10234074,9424952,8771188,10195223}, which rely on fixed formation structures that constrain each pursuer's path to maintain a strict formation. This approach differs from most of our previous work \cite{10251969,9924233,CAO2015150,doi:10.2514/6.2024-0124,doi:10.2514/1.G007539,doi:10.2514/1.G006957}, which, although flexible, did not account for vehicle safety explicitly in changing operating conditions.
        \item  We introduce a novel decision-making protocol for selecting the appropriate pursuer for coordination and designing a collision avoidance strategy for inter-agent safety. This protocol ensures that each pursuer only coordinates with at most one other pursuer when necessary to prevent collisions. Unlike previous approaches \cite{LAN2010381,franchi2016decentralized,10234074,9424952,8771188,10195223} that require extensive coordination and communication between multiple pursuers, our strategy significantly reduces the complexity of control design, preventing the unnecessary maneuvering of multiple vehicles in order to avoid collisions, which may also help in reducing the overall control effort. 
        \item The proposed guidance laws are developed utilizing custom potential function that combines attractive forces with switching repulsive forces to guide the pursuers toward the desired orbit around the target safely and efficiently. This is unlike previous potential function-based target enclosing approaches \cite{doi:10.2514/1.30507,doi:10.2514/1.G002281,1549948,doi:10.1080/00207720902750003} that utilize swarming or flocking-based methods (where repulsive force acts continuously among all neighboring pursuers). Our approach activates repulsive forces only when necessary and only towards one other neighbouring pursuer enabling more independent maneuvers with reduced information.
        \item We demonstrate that our proposed guidance law is robust to the motion of other pursuers in the swarm and requires only relative information. Unlike previous approaches (including some of our previous works \cite{doi:10.2514/1.G007057, CAO2015150}) that rely on global information or information on other pursuers' maneuvers to achieve collision-free target enclosing, our approach potentially eliminates the need for costly agent-to-agent communication and can perform efficiently in environments where global information is not readily available.
    \end{enumerate}         
\section{Problem Formulation}
\subsection{Kinematics of Agents' Relative Motions}
\begin{figure*}[h]
	\centering
	\begin{subfigure}[t]{0.49\linewidth}
		\centering
		\includegraphics[width=0.9\linewidth]{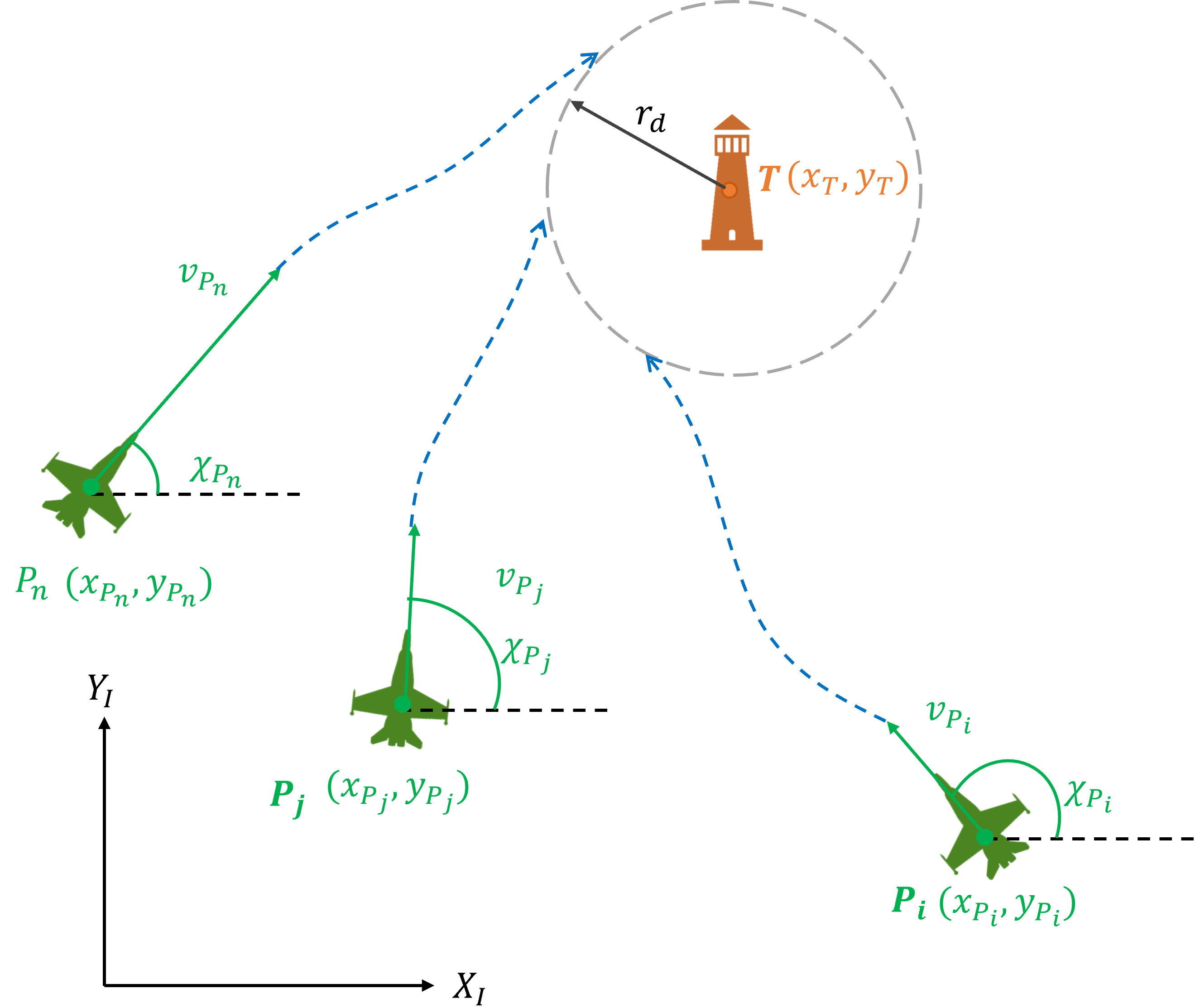}
		\caption{Inertial frame of reference.}
		\label{fig:in_eng}
	\end{subfigure}
	\begin{subfigure}[t]{0.49\linewidth}
		\centering		\includegraphics[width=0.9\linewidth]{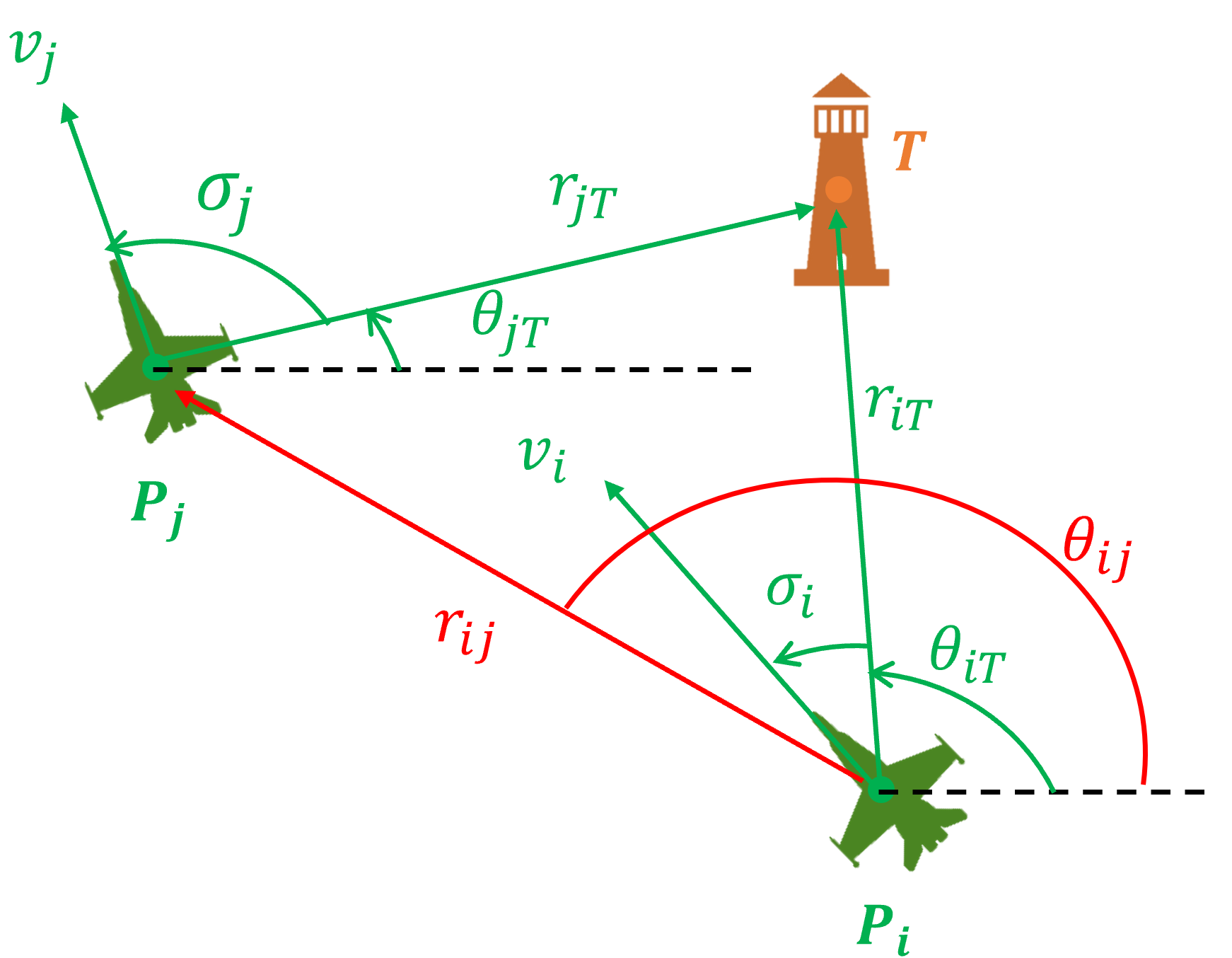}
		\caption{Relative frame of reference.}
		\label{fig:local_eng}
	\end{subfigure}	
 \caption{Multi-agent Target enclosing geometry.}
	\label{fig:tar_enc_sce}
\end{figure*}

Consider a multiagent system consisting of $n$ pursuers and a single stationary target $T$. We use $\mathcal{L}$ to denote the set of the pursuer, that is, $\mathcal{L}=\{P_1,P_2,\ldots,P_n\}$. \Cref{fig:in_eng} depicts the engagement geometry between the pursuers and the target, where 
$[x_i, y_i]^\top \in \mathbb{R}^2$ denotes the position of the $i$\textsuperscript{th} pursuer in the inertial frame of reference, and $\chi_i \in [-\pi, \pi)$ denotes its heading angle. Assuming the vehicles to be point masses, the motion of the $i$\textsuperscript{th} pursuer is governed by 
\begin{align} \notag
    \dot{x}_i=v\cos\chi_i,\quad
     \dot{y}_i=v\sin\chi_i,\quad
     \dot{\chi}_i= \frac{a_i}{v},     
\end{align}
where $v \in \mathbb{R}_{>0}$ denotes the common and constant speed for all pursuers and $a_i$ denotes the lateral acceleration, also the control input, of the $i$\textsuperscript{th} pursuer. This model, albeit simple, is practical when dealing with vehicles that are turn-constrained and utilize lateral forces (e.g., lift and side force) to maneuver, such as fixed-wing aircraft, missiles, and underwater vehicles \cite{doi:10.1080/00207720902750003,10251969,doi:10.2514/1.G007539,9924233,doi:10.2514/6.2024-0124,doi:10.2514/1.G007057}. Here we assume that $\vert a_i\vert\leq a_i^{\max}$.
 

The engagement geometry between the vehicles in the relative frame of reference is depicted in \Cref{fig:local_eng}, where $r_{ij}$ denotes the distance between the $i$\textsuperscript{th} and the $j$\textsuperscript{th} pursuer ($i\neq j$) and accordingly $r_{iT}$ signifies the distance with the target. Similarly, $\theta_{iT}$ denotes the line-of-sight (LOS) angle for the $i$\textsuperscript{th} pursuer with respect to the target and $\theta_{ij}$ is the LOS angle between the $i$\textsuperscript{th} and the $j$\textsuperscript{th} pursuer referenced from the $i$\textsuperscript{th} pursuer.  The bearing angle of the $i$\textsuperscript{th} pursuer with respect to the target is $\sigma_i = \chi_i-\theta_{iT}$. Thus, the kinematics of the relative motion for the multiagent system is given by
\begin{subequations}
\label{eqn:dyn_pur_tar}
\begin{align}
\label{eqn:dyn_pur_tar_a}
    \dot{r}_{iT} &= -v \cos\sigma_i\\
    \label{eqn:dyn_pur_tar_b}
    \dot{\theta}_{iT} & = -\frac{v \sin\sigma_i}{r_{iT}}.
\end{align}
\end{subequations}
These equations are obtained by projecting the $i$\textsuperscript{th} pursuer's relative velocities in the directions parallel/normal to the $i$\textsuperscript{th} pursuer-target LOS. Assuming the pursuers have onboard sensors, we utilize the relative kinematics to develop the control laws minimizing the need for vehicle-to-vehicle communication. Consequently, other pursuers are positioned within each pursuer's sensing range $r_S$ to be detected.

\subsection{Direction/Sense of Target Enclosing}
\label{sec:sirsense}
\begin{figure*}[h]
	\centering
		\includegraphics[width=0.3\linewidth]{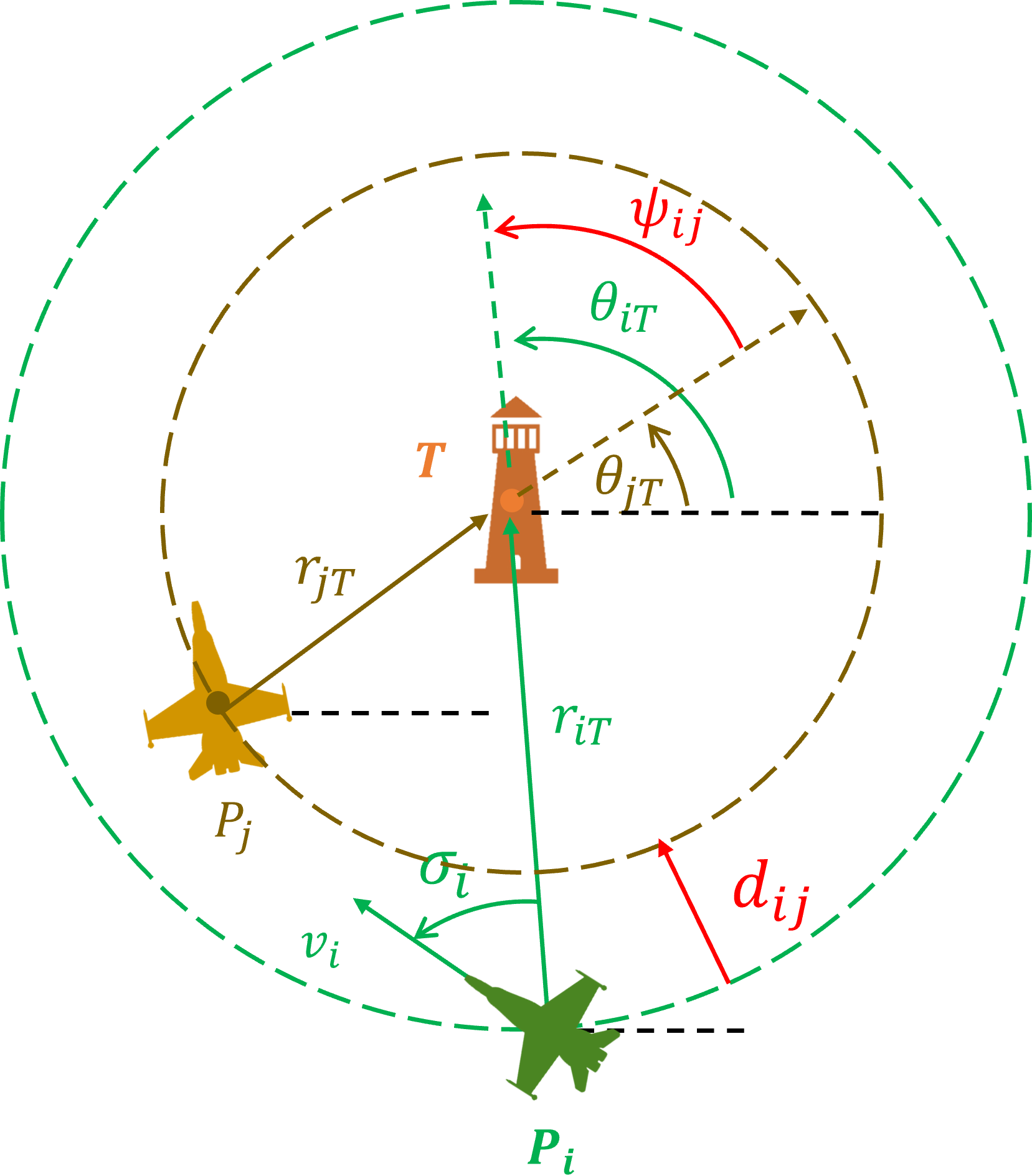}
		\caption{Inter-agent reference geometry.}
		\label{fig:int_pur_eng}
	\label{fig:int_pur}
\end{figure*}

For the target-centric relative system in \eqref{eqn:dyn_pur_tar}, $r_{iT}$ is measured from $P_i$ to target and $\theta_{iT}$ is measured in a counter-clockwise sense from the horizontal reference. Therefore, positive angular rates pertain to the counter-clockwise revolution of the pursuers about the target. Further, it is evident from \eqref{eqn:dyn_pur_tar_b} that $\dot{\theta}_{iT}<0,\; \forall \; \sigma_i \in (0,\pi)$, $\dot{\theta}_{iT}>0,\forall\, \sigma_i \in (-\pi,0)$ and  $\dot{\theta}_{iT}=0, \forall \, \sigma_i \in  \{0, \pm \pi\}$. 
\begin{remark}
One may notice from these observations that the $i$\textsuperscript{th} pursuer revolves around the target in a counter-clockwise direction when $\dot{\theta}_{iT}>0$ and vice-versa. From the bearing angle relationship, it follows that the positive values of $\sigma_i$ represent the scenarios when the $i$\textsuperscript{th} pursuer's heading angle leads the LOS angle in the counter-clockwise direction.
Therefore, if $\sigma_i$ is assumed to be constrained within $(0,\pi)$, then $\dot{\theta}_{iT}<0$ can be ensured for all $t\geq 0$. This aspect is critical from a safety perspective because it limits the $i$\textsuperscript{th} pursuer's revolution to the clockwise direction only around the target. This helps ensure that the pursuers do not go in opposite senses while converging on a circle of the same radius, which could lead to a collision.
\end{remark}
To describe the relationship between the $i$\textsuperscript{th} and the $j$\textsuperscript{th} pursuer at any instant of time in terms of their relative positions, let us define the terms inter-agent separation $d_{ij}$ and the inter-agent angular spacing $\psi_{ij}$ as
 \begin{align}
  \label{eqn:interpur_a}
     d_{ij} = r_{iT}-r_{jT},\;\;
     \psi_{ij} = \theta_{iT}- \theta_{jT},
 \end{align} 
where $\psi_{ij} \in [-\pi, \pi)$, as depicted in \Cref{fig:int_pur_eng}, where the agents (the pursuers) are enclosing the target in a clockwise direction. In \Cref{fig:int_pur_eng}, the dotted circles represent the loiter circles around the target with a radius equal to the instantaneous pursuer-target distance.  The positive values of $d_{ij}$ imply that the $i$\textsuperscript{th} pursuer is on a larger loiter circle compared to the $j$\textsuperscript{th} pursuer. Hence, $d_{ij}>0$ implies $\dfrac{{r}_{iT}}{r_{jT}}>1$, and $d_{ij}<0$ implies that $0<\dfrac{{r}_{iT}}{r_{jT}}<1$. Similarly, the positive values of $\psi_{ij}$ result in a scenario when $\theta_{jT}$ leads $\theta_{iT}$ in a clockwise sense. The red arrows in \Cref{fig:int_pur_eng} denote the positive conventions of $d_{ij}$ and $\psi_{ij}$ considered in this paper. It is also worth noting that for any two agents $P_i$ and $P_j$, $d_{ij}$ and $\psi_{ij}$ together represent the inter-agent radial distance $r_{ij}$. From \Cref{fig:local_eng}, using the cosine angle formula from triangle $P_i-P_j-T$, we can obtain the distance between $P_i$ and $P_j$ as, $r_{ij}^2 =r_{iT}^2 +r_{jT}^2 - 2r_{iT} r_{jT}\cos\psi_{ij}$, which reduces to,
    \begin{align}
       r_{ij}^2 =2r_{jT}^2\left(1-\cos\psi_{ij}\right)+d_{ij}^2 + 2r_{jT}d_{ij}\left(1-\cos\psi_{ij}\right) ,\label{eqn:absdist}
    \end{align} 
    on using $r_{iT}=r_{jT} + d_{ij}$ from \eqref{eqn:interpur_a}. Hence, if both $d_{ij}\to 0$ (\textit{coincident loiter circles}) and $\psi_{ij}\to 0$ (\textit{coincident pursuer-target LOS}), then it essentially implies that $r_{ij}\to 0$, which means inter-agent collision since more than one pursuer occupies the same position.

\subsection{Design of Control Objectives}
We aim to make the pursuers organize themselves in a circular orbit around the target while also ensuring that they do not collide with each other, as the orbit is the same for all of them. To solve the multiagent target enclosing problem under limited information and safety constraints, we need to ensure that for each pursuer, starting from any feasible initial conditions, the following holds:
\begin{align}
    \label{eqn:cond_enclos}
    \lim_{t\to\infty} \left\vert r_{iT}- r_d\right\vert\to 0, 
\end{align}
where $r_d$ denotes the fixed desired proximity of the pursuers from the target, which is the same for every pursuer. It is worth noting that fulfilling the objective in \eqref{eqn:cond_enclos} is not sufficient to guarantee inter-agent safety because the circular orbit is the same for all pursuers during target enclosing. 
In addition to \eqref{eqn:cond_enclos}, the requirement that $r_{ij}>0,\,\forall\,i\neq j,$ needs to be satisfied throughout the engagement for safety. This condition ensures that the inter-agent spacing is always greater than a positive value, that is, no two pursuers converge to the same point on the desired enclosing orbit.
\begin{remark}\label{rmk:rij}
    As the number of agents increases, exhaustively checking the condition of inter-agent spacing for safety ($r_{ij}>0$) leads to computational and design complexities in terms of sensing and communication. For an $i$\textsuperscript{th} agent, there exists some $r_{ij}$ such that checking their sign to ensure safety would be tedious and redundant. That said, an agent may require extensive sensing and communication with every other agent (not just its neighbors), which may not be efficient. 
\end{remark}

\begin{figure*}[h]
	\centering
	\begin{subfigure}[t]{0.49\linewidth}
		\centering
		\includegraphics[width=0.65\linewidth]{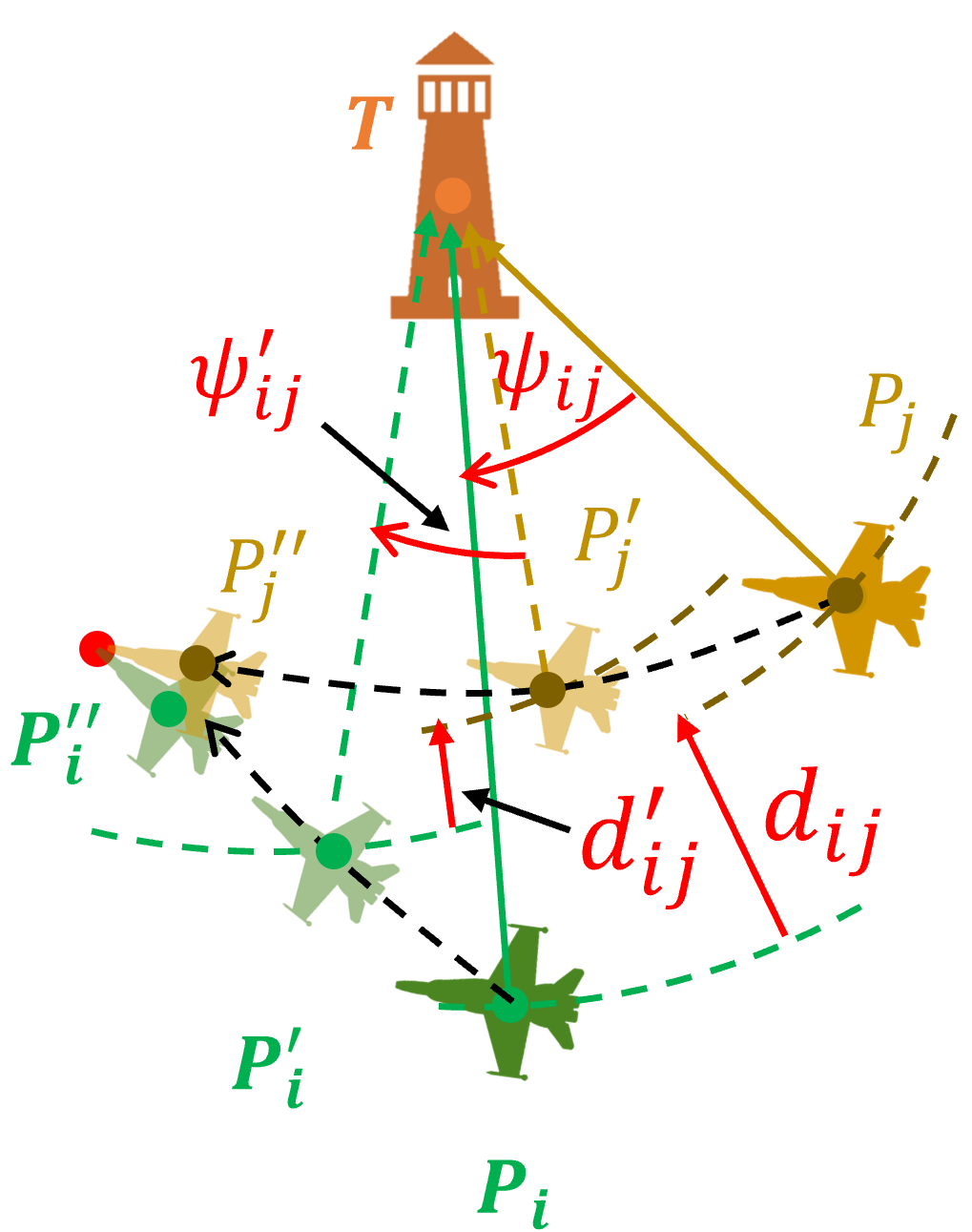}
		\caption{Without safety.}
		\label{fig:int_pur_nearcol}
	\end{subfigure}
	\begin{subfigure}[t]{0.49\linewidth}
		\centering		\includegraphics[width=0.8\linewidth]{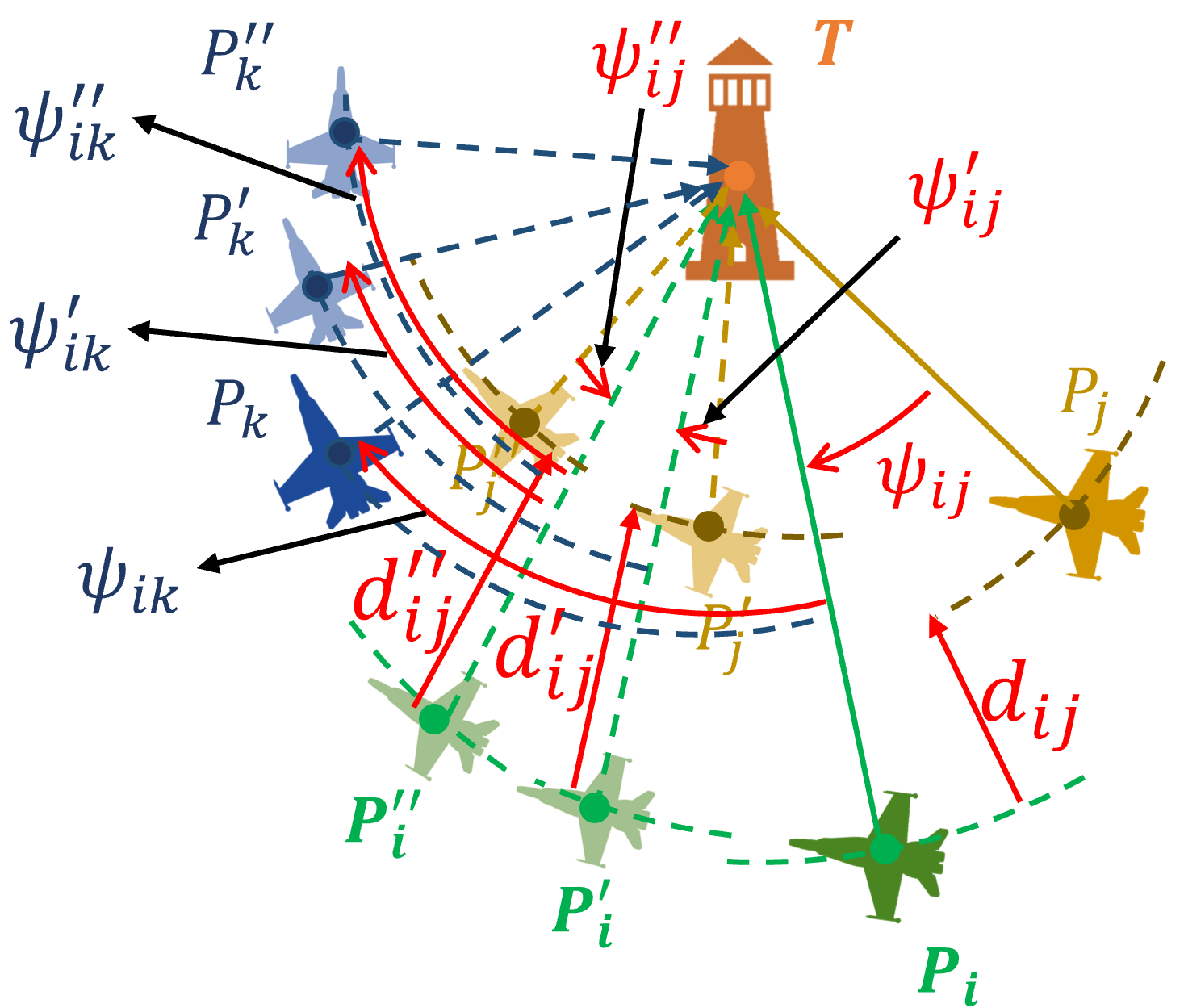}
		\caption{With safety .}
		\label{fig:int_pur_eng_safe}
	\end{subfigure}	
 \caption{Engagement geometry among pursuers in target-enclosing.}
	\label{fig:tar_enc_sce_2}
\end{figure*}
To elucidate the implications of \Cref{rmk:rij}, let us analyze a typical case of target enclosing involving two pursuers, as shown in \Cref{fig:int_pur_nearcol}. If $P_j$ is initially located closer to the target than $P_i$ such that $d_{ij}>0$ and $\psi_{ij}<0$, then $\dot{\theta}_{iT}<\dot{\theta}_{jT}$. This essentially means that the component of the relative velocity of $P_j$ normal to $r_{jT}$ is greater than that of $P_i$, thereby making $\dot{\psi}_{ij}>0$. This leads to coincident pursuer-target LOS between $P_i$ and $P_j$ at some instant of time during target enclosing. Similarly, the component of the relative velocity of $P_j$ along $r_{jT}$ will be less than that of $P_i$, thereby making $\dot{d}_{ij}<0$. This results in the pursuers being on the same loiter circle at some instant of time during target enclosing. Consequently, a collision between $P_i$ and $P_j$ may eventually occur, as demonstrated in \Cref{fig:int_pur_nearcol}, where ${(\cdot)}^{'}$ and ${(\cdot)}^{''}$ represent the variables related to the pursuers as time progresses since both $d_{ij}\to 0$ and $\psi_{ij}\to 0$ as a result of $\dot{d}_{ij}<0$ and $\dot{\psi}_{ij}>0$. Therefore, one may speculate that inter-agent collisions can be prevented if such conditions/scenarios never arise during target enclosing.


The previous analysis showed that collisions between pursuers occur only under specific circumstances. Our method takes advantage of this fact to restrict each pursuer's motion to avoid collisions solely with those pursuers that have the potential to collide in the future. It is important to note that when there are multiple pursuers, several of them may collide with $P_i$. Consequently, we devise strategies to prevent collisions with a single pursuer while taking into account the potential collisions with all the other pursuers. To this end, we first identify the set of all pursuers that are within the sensing range $r_s$ of the $i$\textsuperscript{th} pursuer (i.e., $r_{ij}\leq r_s$) and may potentially collide with it. Let us refer to this set as
\begin{align}
    \mathcal{N}_i \coloneqq \left\{P_j\in \mathcal{L}\,\big\vert\ d_{ij}>0,  \psi_{ij}<0\,\&\,r_{ij}\leq r_s\right\},~\forall\,i\neq j,
\end{align}
which represents the set of all colliding neighbors of $P_i$. Out of all such pursuers in $\mathcal{N}_i$, we identify those pursuers that are situated on a loiter circle whose radius is closest (just smaller) to the one on which $P_i$ is located. Let's define
\begin{align}
\label{eqn:coll_1}
  \mathcal{Z}_i \coloneqq\left\{P_j \in \mathcal{N}_i\big\vert \, d_{ij} =\min\limits_{P_k\in\mathcal{N}_i}d_{ik}\right\}, 
\end{align}
for any $k$\textsuperscript{th} pursuer such that $k\neq j$. Note that the sets $\mathcal{N}_i$ and $\mathcal{Z}_i$ may contain more than one pursuer. This happens when multiple neighboring pursuers belonging to the set $\mathcal{Z}_i$ share the same distance from the target. We, however, aim to develop a self-organizing guidance strategy under limited information and safety constraints. This would entail limiting the communication among the pursuers. Toward this objective, we define another set
\begin{align}
\label{eqn:co_cond}
    \mathcal{C}_i \coloneqq\left\{P_j \in \mathcal{Z}_i\big\vert \,\psi_{ij} =\max\limits_{P_k\in\mathcal{N}_i}\psi_{ik}\right\},
\end{align}
that selects only one potentially colliding neighbor from the set $\mathcal{Z}_i$, thereby allowing us to analyze the behavior of only one pair of pursuers at any instant of time. We refer to the set $\mathcal{C}_i$ as the set of the \textit{nearest colliding pursuer}. In essence, $\mathcal{C}_i\subset \mathcal{Z}_i \subset \mathcal{N}_i$. A pseudocode representation of the algorithm for each pursuer to identify the nearest colliding pursuer is provided below in \Cref{alg:cap}. 
 \begin{algorithm}
            \caption{Select Nearest Colliding Pursuer} \label{alg:cap}
    \label{alg:nearest_colliding_pursuer}
    \begin{algorithmic}[1]
    \State \textbf{Input:} Set of pursuers $\mathcal{L}$, sensing range $r_s$, pursuer $P_i$
    \State \textbf{Output:} Set of nearest colliding pursuer $\mathcal{C}_i$

    \State Initialize $\mathcal{N}_i \gets \{\}$, $\mathcal{Z}_i \gets \{\}$, $\mathcal{C}_i \gets \{\}$
    \For{each pursuer $P_j \in \mathcal{L}$}
        \If{$r_{ij} \leq {r_s}$ and $d_{ij} > 0$ and $\psi_{ij} < 0$}
        \State Add $P_j$ to $\mathcal{N}_i$
        \EndIf
    \EndFor
    \For{each pursuer $P_j \in \mathcal{N}_i$}
        \If{$d_{ij} = \min_{P_k\in\mathcal{N}_i} d_{ik}$}
        \State Add $P_j$ to $\mathcal{Z}_i$
        \EndIf
    \EndFor
    \State Select $P_j \in \mathcal{Z}_i$ such that $\psi_{ij} = \max_{P_k\in\mathcal{N}_i} \psi_{ik}$ 
    \State Add $P_j$ to $\mathcal{C}_i$
    \end{algorithmic}
\end{algorithm}
\begin{remark}
Based on the conditions given in \eqref{eqn:coll_1} and \eqref{eqn:co_cond}, the set $\mathcal{C}_i$ contains at most one neighboring pursuer since a unique combination of $(d_{ij},\psi_{ij})$ represents a distinct position of $P_j$ (the nearest colliding pursuer) about $P_i$. Therefore, it is sufficient to prevent inter-agent collision between any such pair, thereby ensuring safety under limited information.
\end{remark} 
In this work, we show that if $d_{ij}>0$ for all $t\geq 0$ in addition to \eqref{eqn:cond_enclos}, then the pursuers will exhibit a self-organizing behavior during target enclosing while also guaranteeing safety. Ensuring $d_{ij}>0$ will imply that $r_{ij}>0$ even if $\psi_{ij}\to 0$, meaning that every pursuer will maintain a safe distance with respect to every other pursuer in its vicinity. The scenario represented in \Cref{fig:int_pur_eng_safe} considers three pursuers, $P_i$, $P_j$, and $P_k$, aiming to enclose the target within a circle while also ensuring safety. The dotted curves in \Cref{fig:int_pur_eng_safe} represent their instantaneous loiter circles around the target. In this scenario, $P_j$ and $P_k$ never encounter a nearest colliding pursuer before converging to the desired proximity, settling in the same order as they started. Meanwhile, $P_i$ encounters $P_j$ as its nearest colliding pursuer, prompting $P_i$ to maintain $d_{ij}>0$ to avoid collision with $P_j$. Subsequently, at a later time during enclosing, $\psi_{ij}\to 0$, enabling $P_i$ to relax its safety constraint and settle behind $P_j$, ultimately resulting in collision-free self-organizing target enclosing.


This observation facilitates us to formulate the safety constraint for the $i$\textsuperscript{th} pursuer as the one of ensuring 
\begin{align}
    \label{eqn:coll_avoid}
    d_{ij} > 0.
\end{align}
This condition guarantees that all pursuers move along a loiter circle with a radius greater than its distance from the nearest colliding pursuer around the target. As a result, $\psi_{ij} \to 0$, but $d_{ij} \not\to 0$, which implies that $r_{ij} \not\to 0$, ensuring that each pursuer keeps a safe distance from its nearest colliding pursuer and is influenced by at most one other pursuer during enclosing. \Cref{fig:int_pur_eng_safe} shows the behavior of pursuers who ensure safety conditions in \eqref{eqn:coll_avoid}, where $P_j$ is the nearest colliding pursuer for $P_i$. Initially, $\psi_{ij}$ starts at a negative value and becomes positive over time, which is similar to the scenario without safety conditions presented in \Cref{fig:int_pur_nearcol}. However, due to the safety condition, $P_i$ maintains a larger distance from the target than $P_j$, resulting in a positive $d_{ij}$. Once $\psi_{ij}=0$, $P_j$ is no longer the nearest colliding pursuer for $P_i$, and $P_i$ begins to converge to the desired enclosing geometry while $\psi_{ij}$ becomes increasingly positive. At steady state, $P_i$ autonomously converges to a finite positive value of $\psi_{ij}$ with respect to $P_j$. This behavior, where pursuers converge to a circle of desired proximity from the target while maintaining a safe distance from each other, is referred to as self-organizing multiagent target enclosing behavior in our paper. Unlike previous works \cite{KIM20071426,MARSHALL20063,7066874,LAN2010381,franchi2016decentralized,10195223}, the safety condition in \eqref{eqn:coll_avoid} constrains $d_{ij}$ rather than $\psi_{ij}$, allowing each pursuer to independently converge to any finite $\psi_{ij}$ with respect to its neighbors. Such behavior enables pursuers to operate without the need for a predetermined formation structure, reducing the amount of information required and coordinating efforts with just one other pursuer, which is the nearest colliding pursuer. This affords pursuers greater autonomy to maneuver independently while still maintaining a stable formation/enclosing geometry. Moreover, pursuers can enter or exit the enclosing formation seamlessly, without disturbing the overall structure. We are now in a position to summarize the control objectives in regards to the self-organizing multiagent target enclosing under limited information and safety guarantees, that is,
$\lim_{t\to\infty} \left\vert r_{iT}- r_d\right\vert\to 0$, and 
 $d_{ij} > 0,~\forall\,P_j\in\mathcal{C}_i$.
\section{Self-organizing Multiagent Target Enclosing Guidance Law with Inherent Safety}
In this section, we examine the interactions between pursuers and derive the criteria for identifying the nearest colliding pursuer. We then show that satisfaction of the safety condition in \eqref{eqn:coll_avoid} is sufficient for developing a guidance strategy that steers all pursuers on the desired enclosing geometry around the target while also avoiding collisions.
\begin{figure*}[h!]
    \centering \includegraphics[width=0.35\linewidth]{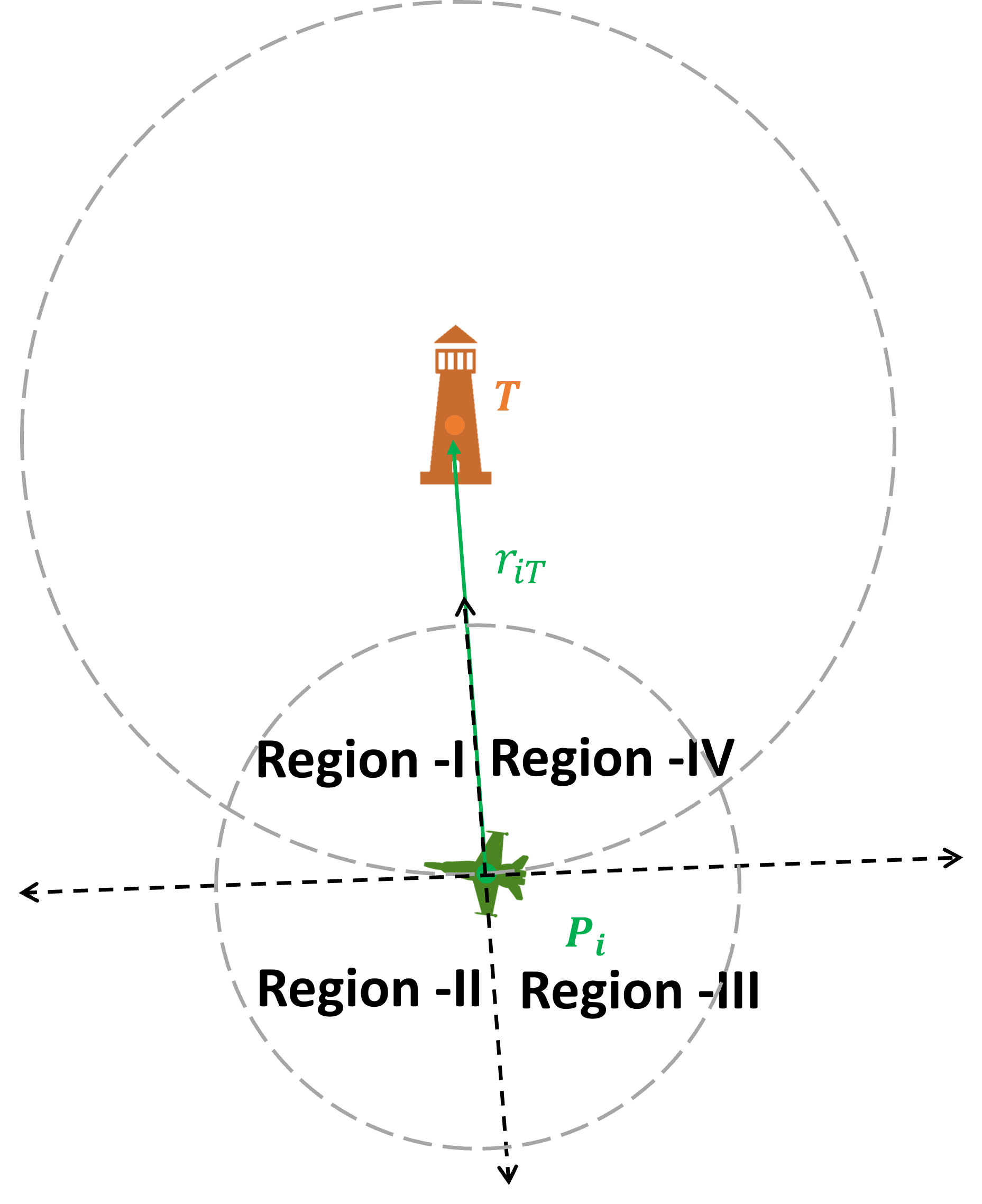}
    \caption{Regions around the $i$\textsuperscript{th} pursuer.}
    \label{fig:proof}
\end{figure*}
\subsection{Analysis of Inter-Pursuer Behavior}
\label{sec:inter_behave}
Note that the LOS of the $i$\textsuperscript{th} pursuer with respect to the target and the line that is normal to it at the $i$\textsuperscript{th} pursuer's position (the tangent line to its instantaneous loiter circle) partitions the engagement zone into four distinct regions that can be completely characterized using $d_{ij}$ and $\psi_{ij}$, as shown in \Cref{fig:proof}. Based on this partition, we can analyze the safety and behaviors of the pursuers as they maneuver to enclose the target. These four distinct region can be characterized as,
\begin{align}
\label{eqn:region}
    \text{Region I:} \; &d_{ij}>0, \psi_{ij}>0, ~~ 
    \text{Region II:} \; &d_{ij}\leq0, \psi_{ij}>0, \nonumber \\
    \text{Region III:} \; &d_{ij}\leq0, \psi_{ij}<0, ~~ 
    \text{Region IV:} \; &d_{ij}>0, \psi_{ij}<0.    
\end{align}
\begin{lemma}\label{lem11}
  Pursuers situated in Regions I and III corresponding to the $i$\textsuperscript{th} pursuer's position cannot collide with the $i$\textsuperscript{th} pursuer, provided all the pursuers start either inside or outside the desired enclosing orbit.
\end{lemma}
\begin{proof}
    To demonstrate this, we examine the motion of neighbouring pursuers $P_j$'s in the four regions as defined in \eqref{eqn:region} of $P_i$ and determine the conditions under which both $d_{ij}\to 0$ and $\psi_{ij}\to 0$ can occur. We do this by deriving the dynamics of the inter-pursuer radial distance and inter-pursuer angular spacing, which are given by,
      \begin{subequations} 
    \label{eqn:interpur_dyn}
    \begin{align}
    \label{eqn:interpur_dyn_a}
     \dot{d}_{ij} &= -v\left(\cos\sigma_i-\cos\sigma_j\right),\\
      \label{eqn:interpur_dyn_b}
     \dot{\psi}_{ij} & =-v\left(\frac{\sin\sigma_i}{r_{iT}}-\frac{\sin\sigma_j}{r_{jT}}\right),
    \end{align}     
    \end{subequations}
    obtained by differentiating \eqref{eqn:interpur_a} with respect to time and using \eqref{eqn:dyn_pur_tar}.  Now, if a collision is imminent, then $d_{ij}\to 0$ and $\psi_{ij}\to 0$. Let us analyze each of these conditions in the four regions to analyze the situation when every pursuer is heading toward the target to enclose it. Using \eqref{eqn:interpur_dyn_a}, we can determine the conditions that lead to $d_{ij}\to 0$ in the four regions as follows:
    \begin{align}
\label{eqn:coincondition}
    \text{Region I:} \: &\dot{d}_{ij}<0, \text{if} \,\sigma_i<\sigma_j, \nonumber \\
     \text{Region II:} \: &\dot{d}_{ij}>0, \text{if} \,\sigma_i>\sigma_j, \nonumber \\
      \text{Region III:} \: &\dot{d}_{ij}>0, \text{if} \,\sigma_i>\sigma_j, \nonumber \\
       \text{Region IV:} \: &\dot{d}_{ij}<0, \text{if} \, \sigma_i<\sigma_j.
\end{align}
Next, we analyze cases when ${\psi}_{ij}\to 0$ when the above conditions for $d_{ij}\to 0$ also hold, provided that all pursuers initiate their movement outside the desired proximity and tend to move toward it to enclose the target. In such a scenario, we can deduce from \eqref{eqn:dyn_pur_tar_a} that $\dot{r}_{iT}<0, \dot{r}_{jT}<0$ (because agents are moving toward the desired orbit regardless of any strategy), thus limiting their bearing angles $\sigma_i,\sigma_j$ within the interval $(0,\pi/2)$. Such a consideration will facilitate easy verification of the signs of $\dot{\psi}_{ij}$ since the sine function is monotonously increasing in  $(0,\pi/2)$. We denote $\sigma_i^\star$ as the look angle about which ${\psi}_{ij}$ changes its sign, given by $\sin\sigma_i^* = \frac{r_{iT}}{r_{jT}} \sin\sigma_j,$
obtained by equating $\dot{\psi}_{ij}$ to zero in \eqref{eqn:interpur_dyn_b}. In Regions I and IV, we have $d_{ij}>0$ or $\dfrac{r_{iT}}{r_{jT}} > 1$, which implies $\sigma_i^* > \sigma_j$. Thus, $\sigma_i < \sigma_j$ leads to $\sin\sigma_i < \dfrac{r_{iT}}{r_{jT}}\sin\sigma_j$, resulting in $\dot{\psi}_{ij} > 0$. Similarly, in Regions II and III, we have $\sigma_i^* < \sigma_j$, and hence $\sigma_i > \sigma_j$ implies $\sin\sigma_i > \dfrac{r_{iT}}{r_{jT}}\sin\sigma_j$, resulting in $\dot{\psi}_{ij} < 0$. Therefore, we can summarize the inferences from the current analysis and those in \eqref{eqn:coincondition} as
\begin{align}
\label{eqn:colincondition}
    \text{Region I:} \: &\dot{d}_{ij}<0,\dot{\psi}_{ij}>0, \text{if} \,\sigma_i<\sigma_j, \nonumber\\
     \text{Region II:} \: &\dot{d}_{ij}>0, \dot{\psi}_{ij}<0, \text{if} \,\sigma_i>\sigma_j, \nonumber \\
      \text{Region III:} \: &\dot{d}_{ij}>0, \dot{\psi}_{ij}<0, \text{if} \,\sigma_i>\sigma_j, \nonumber \\
       \text{Region IV:} \: &\dot{d}_{ij}<0, \dot{\psi}_{ij}>0, \text{if} \,\sigma_i<\sigma_j.
\end{align}
It follows from \eqref{eqn:coincondition} and \eqref{eqn:colincondition} that inter-pursuer collision can only occur in Regions II and IV during target enclosing. Therefore, no agents in Regions I and III will collide with $P_i$. Similarly, we can analyze the signs of $\dot{\psi}_{ij}$ when all the pursuers start inside the desired proximity and move away from the target to reach the enclosing orbit to reach similar conclusions as those stated previously. Note that under such conditions, $\sigma_i,\sigma_j$ must lie in the interval $(\pi/2,\pi)$ to ensure $\dot{r}_{iT}>0$ and $\dot{r}_{jT}>0$. Additionally, the sine of the bearing angles will be strictly decreasing in this interval.
\end{proof}
\begin{remark}
    It is worth noting that \Cref{lem11} does not require exact knowledge of the control law. Instead, it relies on the general behavior of the pursuers that must be exhibited by any strategy designed to enclose the target. In this work, we deliberately exclude scenarios where $P_i$ and its neighbor $P_j$ are positioned in different regions about the desired proximity (some inside and some outside the desired enclosing orbit) for simplicity and an intuitive understanding. Determining the conditions when a collision will not occur in such cases may require the exact knowledge of the control law, which we leave as a future task.
\end{remark}
\Cref{lem11} shows that nearby pursuers in Regions I and III will not collide with $P_i$, so ensuring safety with those pursuers is unnecessary for $P_i$.  This could make the design simpler. Now, we explain why choosing one pursuer from the surrounding pursuers according to \eqref{eqn:co_cond} is necessary to ensure safety among all pursuers.

\begin{theorem}[Inter-pursuer Safety]
    \label{thm:first}
    The selection of a single nearest colliding pursuer according to \eqref{eqn:co_cond} and fulfilling the safety condition in \eqref{eqn:coll_avoid} is sufficient to guarantee collision avoidance during target enclosing.
\end{theorem}
\begin{proof}
   \begin{figure*}[h]
	\centering
	\begin{subfigure}[t]{0.49\linewidth}
		\centering
		\includegraphics[width=0.7\linewidth]{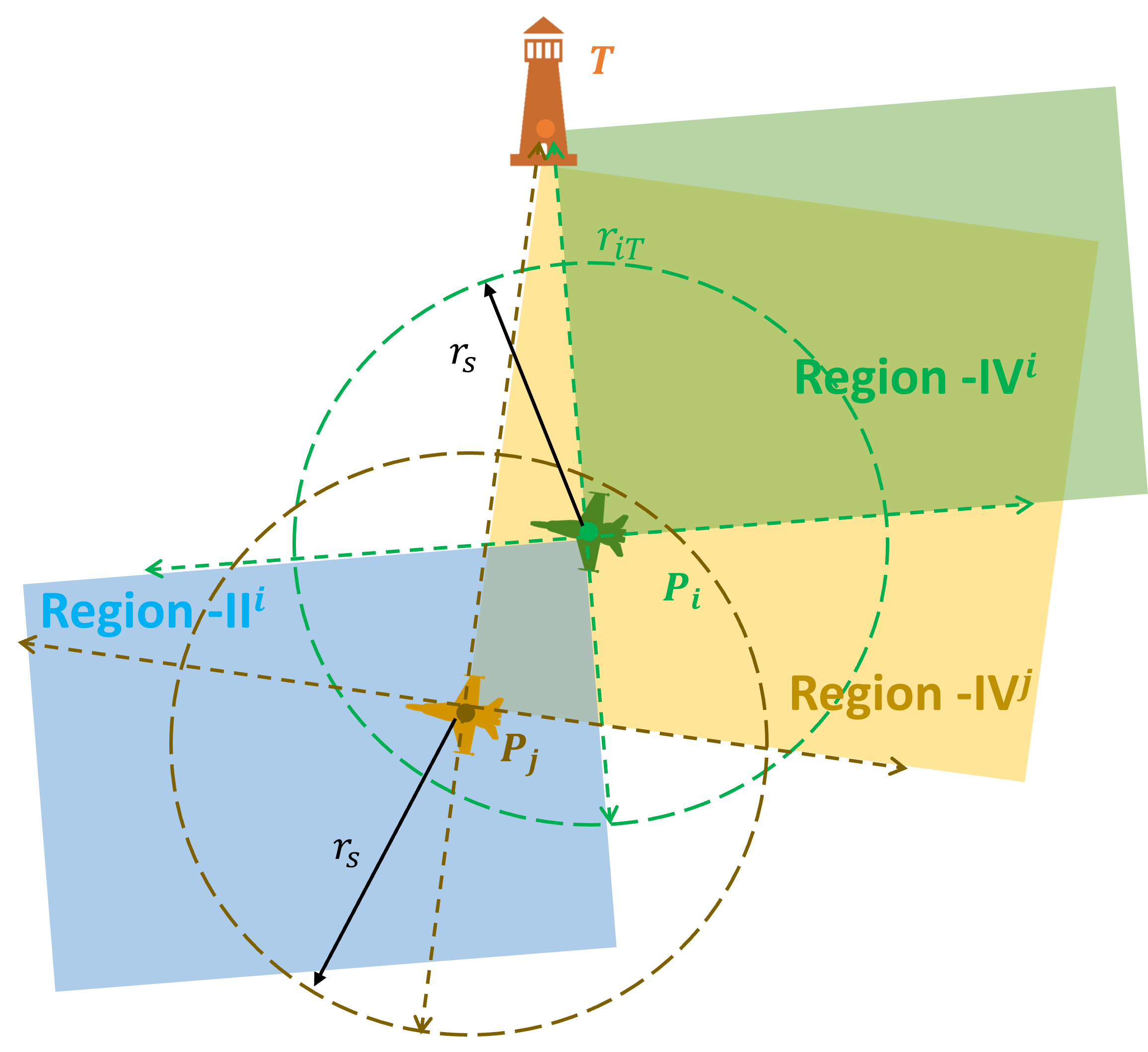}
    \caption{Engagement with neighboring pursuers in Region II of $P_i$.}
    \label{fig:thm_1_proof_1}
	\end{subfigure}
	\begin{subfigure}[t]{0.49\linewidth}
		\centering		\includegraphics[width=0.7\linewidth]{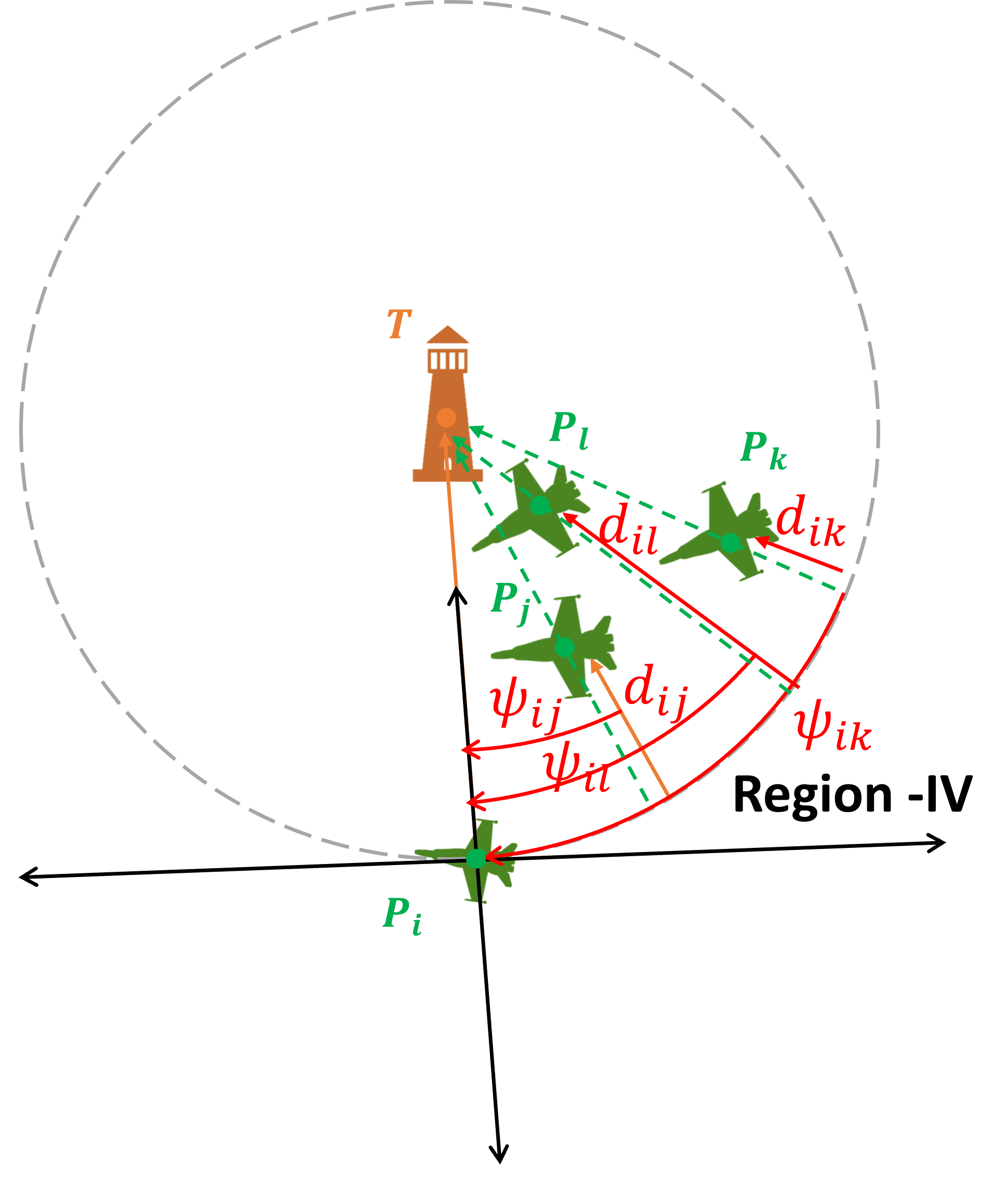}
    \caption{Engagement with neighboring pursuers in Region IV of $P_i$.}
    \label{fig:thm_1_proof_2}
	\end{subfigure}	
 \caption{Engagement in various regions.}
	\label{fig:tar_enc_sce_3}
\end{figure*}
    The essence of this theorem lies in the fact that for any two pursuers, if $d_{ij}>0$, then the $i$\textsuperscript{th} pursuer only needs to ensure strict inter-pursuer safety with its nearest colliding neighbor, which can only be present in its Region II. According to the results in \Cref{lem11}, $P_i$ can only collide with the pursuers located in Regions II and IV, and thus the pursuers located in these regions have a tendency to collide with $P_i$ if safety is not ensured. We show that identifying $P_i$'s nearest colliding neighbor according to \eqref{eqn:co_cond} enables it to selectively choose a single pursuer from a larger group situated in Region IV, thereby enforcing safety guarantees in the self-organizing swarm, that is, $r_{ij}>0$ for $i\neq j$.


    Let us consider a situation where pursuer $P_i$ has another pursuer, $P_j$, located within its Region II (denoted as Region II\textsuperscript{i}), as shown in \Cref{fig:thm_1_proof_1}. Since $P_i$ is the nearest colliding pursuer for $P_j$, according to \eqref{eqn:co_cond}, $P_j$ must meet the safety criterion in \eqref{eqn:coll_avoid} and maneuver accordingly. Hence, $P_i$ can safely approach the desired proximity without colliding with pursuers in its Region II\textsuperscript{i}. This means that $P_i$ can focus on avoiding collisions with pursuers in its Region IV\textsuperscript{i}.
    
    Before considering the second scenario, we first analyze the inter-pursuer distance $r_{ij}$ between $P_i$ and $P_j$ to arrive at the nearest colliding pursuer selection criterion. To find the values of $d_{ij}$ and $\psi_{ij}$ that minimizes $r_{ij}$ by evaluating the partial derivatives of $r_{ij}^2$ as in \eqref{eqn:absdist} with respect to $d_{ij}$ and $\psi_{ij}$. We obtain the gradient $\nabla r_{ij}^2$ and Hessian $\nabla^2 r_{ij}^2$ as,
    $ \nabla r_{ij}^2=
    \begin{bmatrix}
        \dfrac{\partial r_{ij}^2}{\partial d_{ij}} & \dfrac{\partial r_{ij}^2}{\partial \psi_{ij}}
    \end{bmatrix}^\top=\begin{bmatrix}
        2d_{ij} + 2r_{iT}\left(1-\cos\psi_{ij}\right) \\ 2r_{iT}^2\sin\psi_{ij}+2r_{iT}d_{ij}\sin\psi_{ij} 
    \end{bmatrix}$, 
    $ \nabla^2 r_{ij}^2=
    \begin{bmatrix}
        2 & 2r_{iT}\sin\psi_{ij}\\ 2r_{iT}\sin\psi_{ij} &  2r_{iT}^2\cos\psi_{ij}+2r_{iT}d_{ij}\cos\psi_{ij}
    \end{bmatrix}$. It can be readily observed that the minimum value of $r_{ij}=0$ is attained only when $d_{ij}=0$ and $\psi_{ij}=0$, since at these values $\nabla r_{ij}^2=0$ and $\nabla^2 r_{ij}^2 \succ  0$ (positive definite). 
    These inferences follow the discussions in \Cref{sec:sirsense}. Therefore, minimizing both $d_{ij}$ and $\psi_{ij}$ will single out a pursuer with minimum distance $r_{ij}$ from $P_i$. However, such a pursuer might not be suitable for collision avoidance mechanism ensuring $d_{ij}>0$ as in \eqref{eqn:coll_avoid}. Instead, our approach selects the nearest colliding pursuer as per \Cref{alg:cap} to ensure safety among all the pursuers. We know from \eqref{eqn:region} that in Region IV of $P_i$ (Region IV\textsuperscript{i} in \Cref{fig:thm_1_proof_1}), $d_{ik}>0$ and $\psi_{ik}<0$ for an arbitrary pursuer $P_k$ located therein. Therefore, when $r_{ik} \neq 0$, choosing $P_k$ with the smallest $d_{ik}$ and largest $\psi_{ik}$ will select only one pursuer that is closest to $P_i$ in terms of inter-pursuer distance. Our approach ensures safety among pursuers allowing them to self-organize during target enclosing by first picking pursuers located on the nearest loiter circle around the target to minimize $d_{ij}$. This set of pursuers is denoted by $\mathcal{Z}_i$, as seen in \eqref{eqn:coll_1}. Next, we select the nearest colliding pursuer with the largest $\psi_{ij}$ from the set $\mathcal{Z}_i$, fulfilling the conditions in \eqref{eqn:co_cond}. 

In the second situation, we consider the presence of multiple pursuers ($P_j$, $P_k$ and $P_l$) in Region IV\textsuperscript{i}, as shown in \Cref{fig:thm_1_proof_2}. Given the scenario, pursuer $P_j$ has the shortest distance to $P_i$ ($r_{ij}$), while pursuer $P_k$ has the instantaneous loiter circle closest to $P_i$'s instantaneous loiter circle or minimum $d_{ik}$. Additionally, pursuer $P_l$ has the largest inter-pursuer angular spacing between itself and $P_i$ ($\psi_{il}$). This implies $d_{ik}<d_{ij}<d_{il}$, and $\psi_{il}>\psi_{ij}>\psi_{ik}$. From \eqref{eqn:co_cond}, we can infer that $P_k$ is the nearest colliding pursuer for $P_i$, who only needs to concern itself with avoiding collision with $P_k$ by executing a suitable maneuver.

For the sake of contradiction, let us consider either $P_j$ or $P_l$ to be the nearest colliding pursuer for $P_i$. First, considering $P_j$ to be the nearest colliding pursuer for $P_i$, we have $d_{ij}>0$ based on condition \eqref{eqn:coll_avoid}, which also implies $d_{il}>0$. However, this will not ensure  $d_{ik}>0$ and result in a collision between $P_i$ and $P_k$. Similarly, we can show that selecting $P_l$ as the nearest colliding pursuer will not ensure the prevention of collision of $P_i$ with $P_j$ and $P_k$, since $d_{il}>0$ does not imply $d_{ij}>0$ and $d_{ik}>0$. Therefore, $P_k$ is the appropriate choice of pursuer chosen as per condition \eqref{eqn:co_cond}, which requires less information than ensuring safety with a pursuer other than the nearest colliding one (which will require additional conditions).  By guaranteeing safety condition \eqref{eqn:coll_avoid} with respect to $P_k$, we can infer that $d_{ik}>0$. This subsequently implies that $d_{ij}>0$ and $d_{il}>0$, since $P_k$ is closer to $P_i$ than $P_j$ and $P_l$. In other words, ensuring that $P_i$ avoids colliding with $P_k$ automatically ensures that $P_i$ avoids colliding with $P_j$ and $P_l$ as well. This concludes the proof.
\end{proof}
Every pursuer starting away from the target may encounter a nearest colliding pursuer before reaching the desired proximity around the target. Under such a case, the $i$\textsuperscript{th} pursuer should maneuver to only avoid collision with its nearest colliding pursuer, ensuring safety conditions in \eqref{eqn:coll_avoid}. Consequently, this allows the pursuers to settle behind the nearest colliding pursuer on the enclosing orbit, allowing the pursuers to self-organize around the target with inherent safety. \begin{remark}
    Our methodology capitalizes on the inherent self-organization behavior of each pursuer to adjust their individual trajectory without necessitating explicit cooperation or prior coordination among them. This decentralized approach allows pursuers to operate independently, adapting to changing environmental conditions and avoiding potential collisions while still working towards the collective goal of enclosing the target. By eliminating the need for rigid formations or vehicle-to-vehicle communication, our solution offers a more practical and robust alternative to traditional centralized control strategies.
\end{remark}

\label{secIII}
\subsection{Design of Potential Function}
\label{potentialfunc}

As indicated by the behavior of pursuers observed in \Cref{sec:inter_behave}, each pursuer must exhibit two distinct behaviors. When there is no nearby colliding pursuer, the $i$\textsuperscript{th} pursuer's objective is to reach the desired proximity from the target to enclose it. However, if there is a nearest colliding pursuer, the $i$\textsuperscript{th} pursuer needs to maneuver along a loiter circle with a radius larger than the one corresponding to the nearest colliding pursuer to prevent any potential collisions. These factors enable us to break down the complex multiagent target enclosing problem into more manageable two-body (one pursuer and the target) or three-body (two pursuers and the target) scenarios at any given point in time. Consequently, in this subsection, we present the design of a potential function that will serve as the foundation for developing the target enclosing guidance law.
 
Let's define the range error for the $i$\textsuperscript{th} pursuer as, ${e}_i = r_{iT} - r_d,$
where $r_d$ is the desired fixed proximity from the target such that $\dot{r}_d=0$. This error variable plays a crucial role in the design of the potential function to ensure inter-pursuer safety. To analyze the error dynamics, we differentiate $e_i$ with respect to time and use \eqref{eqn:dyn_pur_tar_a} to obtain the range error rate for the $i$\textsuperscript{th} pursuer as, $\dot{e}_i = \dot{r}_{iT}=-v\cos\sigma_i$,
since $\dot{r}_d=0$. From the above equation, it is evident that $\dot{e}_i<0,$ if $ \sigma_i \in [0,\pi/2)$, which indicates the $i$\textsuperscript{th} pursuer's motion toward the target, whereas $\dot{e}_i>0,$ represents the opposite case because $\sigma_i \in (\pi/2,\pi]$. The $i$\textsuperscript{th} pursuer moves on a circle of fixed radius from the target when $\dot{e}_i=0$ when $\sigma_i= \pi/2$. Therefore, nullifying the range error $e_i$ will ensure that the $i$\textsuperscript{th} pursuer converges to the desired proximity around the target, but additional safety considerations are needed to prevent the inter-pursuer collision.

We propose a potential function for the $i$\textsuperscript{th} pursuer that incorporates both the required behaviors, which is defined as 
\begin{align}
\label{eqn:potent_func}
    \mathcal{U}_i(e_i,e_j) = \lambda_i \frac{e_i^2}{2} + \eta_i \delta_{i} e^{-\frac{(e_i-e_j)^2}{2\Delta_i^2}}
\end{align}
where $\lambda_i$, $\eta_i$, and $\Delta_i$ are positive scaling constants in the potential function. The subscript $j$ corresponds to $P_j \in\mathcal{C}_i$, which is the nearest colliding pursuer corresponding to the $i$\textsuperscript{th} pursuer. The variable $\delta_{i}$ is a conditional one such that $ \delta_{i} =\begin{cases}1; \,&\text{if} \,\mathcal{C}_i \neq\emptyset~(\mathcal{C}_i~\text{is not empty}), \\      
    0;\; &\text{otherwise}.
    \end{cases}$.
The first term in \eqref{eqn:potent_func} is the attractive component of the potential function, which is active at all times. This term has a unique minimum at $e_i=0$ and steers the $i$\textsuperscript{th} pursuer to the desired proximity around the target. The second term in \eqref{eqn:potent_func} is the repulsive term enabling the $i$\textsuperscript{th} pursuer to avoid collisions with its nearest colliding pursuer. It is important to note that the latter term is conditional and is only activated based on the existence of the nearest colliding pursuer for $P_i$.  The first term in \eqref{eqn:potent_func} is a quadratic term dependent on $e_i$, whereas the second term (which introduces a repulsive behavior) is a Gaussian function that attains maximum values at $e_i=e_j$ and decreases as $d_{ij}=e_i-e_j$ increases, with this term losing its dominance when $d_{ij}\geq \Delta_i$. The Gaussian function in \eqref{eqn:potent_func} is chosen as the potential function for collision avoidance due to its localized influence, characterized by an exponential decrease in value with distance from the mean. This function's shape can be precisely controlled through its parameters, such as the mean and standard deviation, offering significant adaptability. In comparison, alternative repulsive potential functions, such as inverse proportional or logarithmic functions, exhibit non-smooth behavior, reduced adaptability, and a rapid decrease to zero as the distance from the mean increases. These attributes may render them less suitable for effective collision avoidance in a dynamically changing environment. If two agents are closer to each other, it is reasonable to assume $d_{ij}<e_i$. Therefore, the first term in \eqref{eqn:potent_func} will dominate the second term for any selected value of $\lambda_i$, $\eta_i$ and $\Delta_i$ if $d_{ij}\gg 0$. However, for small values of $d_{ij}$, one can select the values of these parameters to ensure that the second term dominates in the region where $d_{ij}=e_i-e_j \in (0,\Delta_i)$. In such a scenario, in the region near the line $e_i-e_j=0$, the repulsive term dominates the attractive component that will result in $P_i$ moving away from $P_j \in \mathcal{C}_i$. Consequently, it follows that there will be a shift in the minimum of the potential function at $e_i=0$ (for only attractive potential) to $e_i>e_j$ (for combined potential), which allows pursuers to avoid collisions with each other.

\begin{figure*}[h]
	\centering
	\begin{subfigure}[t]{0.49\linewidth}
		\centering
		\includegraphics[width=\linewidth]{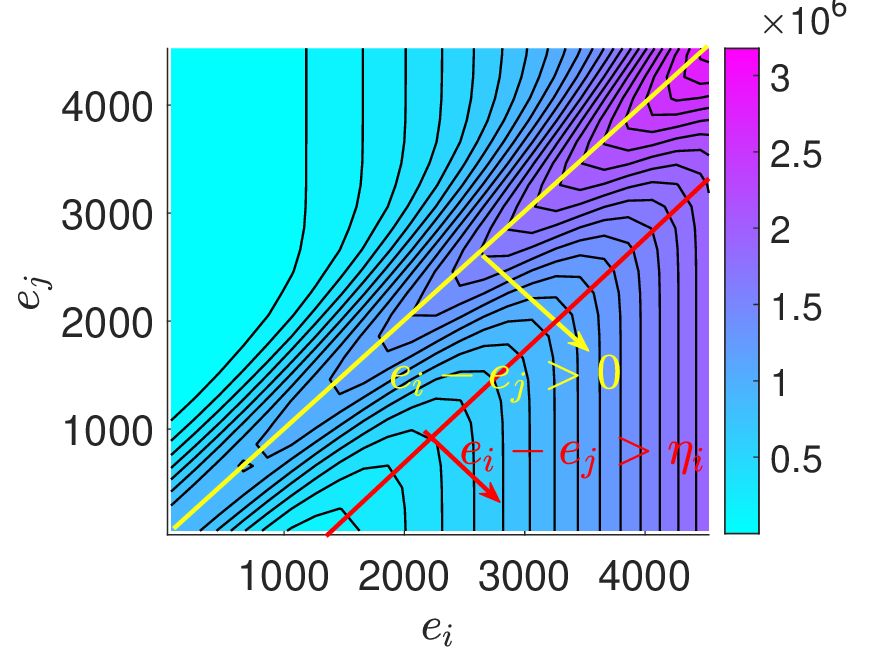}
		\caption{Potential function $\mathcal{U}_i$.}
		\label{fig:potent_1}
	\end{subfigure}
	\begin{subfigure}[t]{0.49\linewidth}
		\centering
		\includegraphics[width=\linewidth]{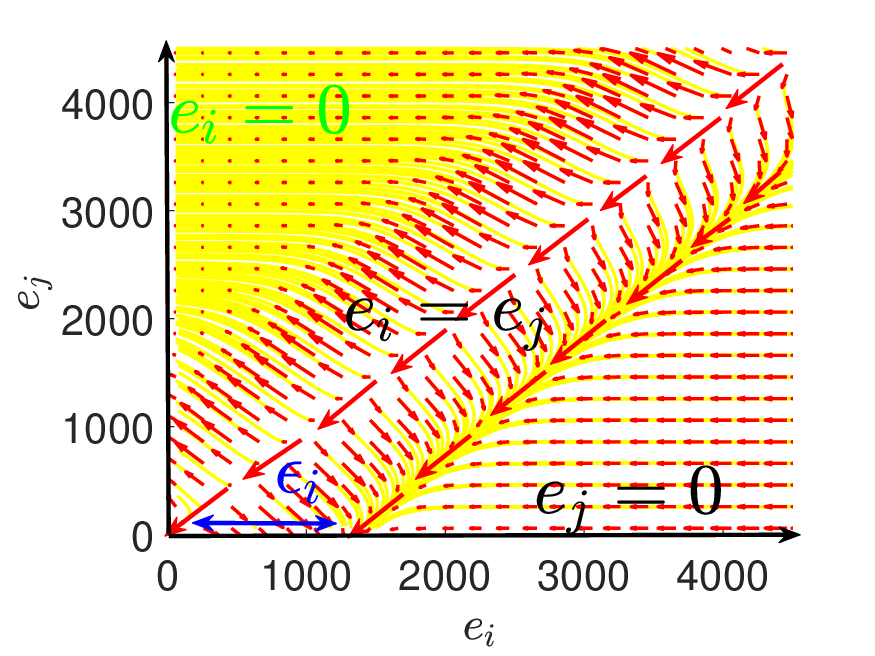}
		\caption{Negative Gradient lines.}
		\label{fig:grad_line}
	\end{subfigure}	
	\caption{Potential function and gradient lines for the $P_i$ in presence of a nearest colliding pursuer $P_j$, with $\lambda_i=0.2, \eta_i=100000$ and $ \Delta_i=500$.}
	\label{fig:potent}
\end{figure*}

To analyze the combined potential function, we compute partial derivatives of \eqref{eqn:potent_func} with respect to $e_i$ and $e_j$ to obtain the gradient of the potential function as,
\begin{subequations}
\label{eqn:grad}
\begin{align}
    \label{grad_1}
    \frac{\partial \mathcal{U}_{i}}{\partial e_i} &=\lambda_i e_i - \frac{\eta_i \delta_{i}}{\Delta^2}e^{-\frac{(e_i-e_j)^2}{2\Delta_i^2}}(e_i-e_j),\\
     \label{grad_2}
    \frac{\partial {\mathcal{U}_{i}}}{\partial e_j} &=  \frac{\eta_i \delta_{i}}{\Delta^2}e^{-\frac{(e_i-e_j)^2}{2\Delta_i^2}}(e_i-e_j).
\end{align}    
\end{subequations}
To obtain the region of minimum potential, we equate \eqref{grad_1} to zero to obtain, that is, $\lambda_i e_i - \frac{\eta_i \delta_{i}}{\Delta^2}e^{-\frac{(e_i-e_j)^2}{2\Delta_i^2}}(e_i-e_j)=0.$
This above expression is highly nonlinear, and the analytical solution of $e_i$ is difficult to obtain in closed form. The trivial solution is $e_i = 0$ and $e_j = 0$. However, such a situation never arises because when $\delta_{i}=1$, $e_i>e_j$. Further, from the above equation, we can infer that the solution to $\dfrac{\partial {\mathcal{U}_{i}}}{\partial e_i}=0$ and $\dfrac{\partial {\mathcal{U}_{i}}}{\partial e_j}=0$, can be given by $e_i-e_j=\epsilon_i$, where $\epsilon_i$ denotes a positive value. To verify this, we plot the potential function and the negative gradients for the chosen (typical) values of $\lambda_i=0.2, \eta_i=100000$ and $ \Delta_i=500$, as shown in \Cref{fig:potent}. In \Cref{fig:potent}, $e_i-e_j>0$ is the relevant area whenever $\delta_i=1$. The combined potential is depicted in \Cref{fig:potent_1}, where it is observed that the minimum potential region (depicted by cyan colored region) is situated at $e_i-e_j=\epsilon_i$. Further, in \Cref{fig:grad_line}, the negative gradient lines converge to the line $e_i-e_j=\epsilon_i$. This depicts the existence of a minimum potential point at $e_i=e_j+\epsilon_i$, when the repulsive component of the potential is activated.   
\begin{remark}
    \label{repl_rem}
    It is crucial to choose the values of the control parameters $\lambda_i$, $\eta_i$, and $\Delta_i$ judiciously to guarantee that $\epsilon_i>0$. This ensures that the minimum of the combined potential function in \eqref{eqn:potent_func} occurs at $e_i>e_j$, leading to $\dot{e}_i>0$ for all $e_i-e_j \in (0,\epsilon_i)$ and $\dot{e}_i<0$ for all $e_i-e_j \in (\epsilon_i,\infty)$. {To ensure that the repulsive term in potential function gradient in \eqref{grad_1} dominates whenever activated, the condition $\dfrac{\delta_i}{\Delta_i^2\lambda_i}>\sup_{t\geq 0}\frac{e_i}{e_i-e_j}\exp\left\{\left(\frac{\left(e_i-e_J\right)^2}{2\Delta_i^2}\right)\right\}$ is obtained, which will help select the control parameters to ensure $\epsilon_i>0$. }
\end{remark}


\subsection{Development of the Target Enclosing Guidance Law}
In this section, we design the lateral acceleration (control input) for the $i$\textsuperscript{th} pursuer to steer it on a collision-free path to the desired proximity around the target. To achieve this, we leverage the sliding mode control avowed for its remarkable properties of precision, robustness, and ease of designing, e.g., see \cite{174678,1549948,10251969,doi:10.2514/6.2024-0124,doi:10.2514/1.G007539,doi:10.2514/1.G006957} .
\begin{lemma}\label{lem:reldeg}
    The dynamics of the $i$\textsuperscript{th} pursuer's range error has a relative degree of two with respect to its lateral acceleration.
\end{lemma}
\begin{proof}
    Differentiating $\dot{e}_i$ with respect to time and using \eqref{eqn:dyn_pur_tar_b}, we obtain the dynamics of range error rate as, $\ddot{e}_i = \ddot{r}_{iT} = \dfrac{v^2 \sin^2{\sigma_i}}{r_{iT}} + a_i \sin\sigma_i,$ thereby completing the proof.
\end{proof}
Inferences from \Cref{lem:reldeg} allow us to construct a suitable sliding manifold for the $i$\textsuperscript{th} pursuer. To this end, consider the sliding manifold,
\begin{align}
\label{eqn:slid_final}
    S_i={\dot{e}}_i + \lambda_i {e}_i - \frac{\eta_i \delta_{i}}{\Delta_i^2}\exp\left\{{-\frac{({e}_i-{e}_j)^2}{2\Delta_i^2}}\right\}({e}_i-{e}_j),
\end{align}
on which the $i$\textsuperscript{th} pursuer's range error rate $\dot{e}_i$ can be driven to the negative gradient $-\nabla_{e_i}{\mathcal{U}_{i}}$.
Note from the above equation that when $\delta_i=0$ $S_i=\dot{e}_i+\lambda_ie_i$, which is analogous to a second-order sliding manifold with asymptotic error convergence. Further, the last term in \eqref{eqn:slid_final} represents the collision avoidance term that becomes active when $\delta_i=1$. We now propose the lateral acceleration (the only control input) for the $i$\textsuperscript{th} pursuer as,
\begin{align}
    \label{eqn:ctrl}
    a_i=
    &\dfrac{1}{\sin\sigma_i}\Bigg[-K_i\sign({S_i})- \frac{v^2 \sin^2{\sigma_i}}{r_{iT}} +\lambda_iv\cos\sigma_i \nonumber \\&+ \frac{\delta_{i} \eta_i v e^{-\frac{({e}_i-{e}_j)^2}{2\sigma^2}}}{\Delta_i^2}\left(\frac{({e}_i-{e}_j)^2}{\Delta_i^2}-1\right) \cos\sigma_i\Bigg],
\end{align}
where $K_i > \frac{\eta_i v}{\Delta_i^2}\left(1-\frac{{r_s}^2}{\Delta_i^2}\right)$ provides a sufficient condition on the controller gain of the $i$\textsuperscript{th} pursuer. The first term in the above equation drives the pursuers to their respective sliding manifold. The second term represents the centripetal acceleration about the target, while the third term represents the component of its velocity along its LOS with respect to the target. The fourth term in \eqref{eqn:ctrl}  enforces a collision avoidance behavior. 
\begin{theorem}
\label{thm1}
    Consider the equations of relative motion between a single target and multiple pursuers \eqref{eqn:dyn_pur_tar}, and the $i$\textsuperscript{th} pursuer-target range error $e_i$. If the $i$\textsuperscript{th} pursuer applies the lateral acceleration given in \eqref{eqn:ctrl}, then all the pursuers will self-organize on {a} circle of radius $r_d$ around target without colliding with each other.
\end{theorem}
\begin{proof}
Consider a Lyapunov Function Candidate 
        $V = \sum_{i\in\mathcal{L}}{\frac{1}{2}}S_i^2.$
Differentiating this Lyapunov function with respect to time and using \eqref{eqn:slid_final}, we obtain the dynamics of the sliding manifold as, $\dot{V} = \sum_{i\in\mathcal{L}} S_i \dot{S}_i=\sum_{i\in\mathcal{L}} S_i\left(\ddot{e}_{i} +\lambda_i \dot{e}_i + \frac{\partial^2 \mathcal{U}_{i}}{\partial^2 {e}_i}\dot{e}_i+  \frac{\partial^2 \mathcal{U}_{i}}{\partial^2 {e}_j}\dot{e}_j\right).$
Substituting for $\ddot{e}_i, \,\partial^2\mathcal{U}_i/\partial^2e_i$ and  $\partial^2\mathcal{U}_i/\partial^2e_j$ in the above expression we obtain
$
        \dot{V} =\sum_{i\in\mathcal{L}} S_i \Biggr[\frac{v^2 \sin^2{\sigma_i}}{r_{iT}}+a_i\sin\sigma_i +\lambda\dot{e}_i
         +\frac{\eta_i\delta_{i}e^{-\frac{({e}_i-{e}_j)^2}{2\Delta^2}}}{\Delta_i^2}\left(\frac{({e}_i-{e}_j)^2}{\Delta_i^2}-1\right)\left(\dot{e}_i -\dot{e}_j\right)\Biggr].
$
    Substituting the proposed lateral acceleration for the $i$\textsuperscript{th} pursuer given in \eqref{eqn:ctrl}, reduces the above expression to
    \begin{align}
        \dot{V}&=\sum_{j\in\mathcal{L}}S_i\Bigg[-K_i\sign(S_i) -\frac{\eta_i\delta_{i} e^{-\frac{({e}_i-{e}_j)^2}{2\Delta^2}}}{\Delta_i^2}\left(\frac{({e}_i-{e}_j)^2}{\Delta_i^2}-1\right)\dot{e}_j\Bigg], \nonumber\\
        &\leq\sum_{j \in \mathcal{L}}-\vert S_i\vert\Bigg[K_i-\frac{\eta_i}{\Delta_i^2}\bigg| e^{-\frac{({e}_i-{e}_j)^2}{2\Delta^2}}\bigg|\, \vert\dot{e}_j\vert \,\Bigg|\frac{({e}_i-{e}_j)^2}{\Delta^2}-1\Bigg|\Bigg], \nonumber
        \end{align}
    since $\vert\dot{e}_j\vert=v$ and $\vert e_i-e_j \vert<r_s$ (as the switching term can be activated only when the colliding vehicle is within sensing range). The above equation can be further simplified as, $\dot{V}\leq \sum_{j \in \mathcal{L}}-\vert S_i\vert \Bigg[K_i-\frac{\eta_i v}{\Delta_i^2}\left(1-\frac{{r_s}^2}{\Delta_i^2}\right)\Bigg],$
    which presents the sufficient condition as given after \eqref{eqn:ctrl} to ensure,  $\dot{V}<0, \, \forall \,S_i \in \mathbb{R}\setminus \{0\}$. This implies that $S_i$'s converge to zero within finite time given by $ T^* =\vert S_i(0)\vert/K_i$, irrespective whether $\delta_i=0\,\textup{or}\,1$. It is worth mentioning that $S_i$ can suddenly deviate from zero due to the activation/deactivation of the repulsive potential component. However, based on our control law, even when $S_i$ shifts suddenly to a non-zero value, $S_i$ will converge to a zero within a finite time to quickly enforce the sliding mode. Once $S_i=0$, the pursuers'  error rate propagates according to ${\dot{e}}_i= - \lambda_i {e}_i + \frac{\eta_i \delta_{i}}{\Delta_i^2}e^{-\frac{({e}_i-{e}_j)^2}{2\Delta_i^2}}({e}_i-{e}_j)=0.$
    Therefore, once the sliding mode is enforced, $P_i$ may exhibit two distinct behaviors in relation to other vehicles based on the switching term, discussed as Modes 1 and 2 below. 
    
    \textbf{Mode 1:} When $ \delta_{i} = 0$, indicating the absence of any nearest colliding pursuer for $P_i$, then the error dynamics is described as ${\dot{e}}_i=-\lambda_i {e}_i$ (since $\delta_i=0$). This leads to  1) ${\dot{e}}_i\leq 0$ when $e_i\geq 0$ and  2) ${\dot{e}}_i > 0$ when $e_i<0$. The solution of the range error can obtained as $\; {e}_i(t)={e}_i(0)e^{-\lambda_i t}$. Consequently, the pursuer converges to the desired proximity, that is, $r_{iT} \rightarrow r_d$ if $\lambda_i$ is selected as a positive value.
    
    \textbf{Mode II:} If ${\delta}_{i}=1$, this pertains to the scenario when there exists a nearest colliding neighbor $P_j$ for the pursuer $P_i$ such that $d_{ij}>0$ and $\psi_{ij}<0$ at the start of the engagement. the error dynamics is described as ${\dot{e}}_i= - \lambda_i {e}_i + \frac{\eta_i }{\Delta^2}e^{-\frac{({e}_i-{e}_j)^2}{2\Delta^2}}({e}_i-{e}_j)=0,$
    as $\delta_i=1$. It can be inferred that the solution of $e_i(t)$ when $\delta_i=1$ may not be feasible to obtain analytically in each case. However, we can select the values of controller parameters $\lambda_i$, $\eta_i$ and $\Delta_i$ to ensure $\dot{e_i}>0$ when $e_i-e_j<\epsilon_i$ and $\dot{e_i}<0$ when $e_i-e_j>\epsilon_i$, where $\epsilon_i>0$ denotes a positive value. This is similar to the behavior of the pursuer under the proposed potential function as discussed earlier in \Cref{potentialfunc}. This implies $P_i$ will converge to a loiter circle of radius greater than that of $P_j$, that is, $e_i\to e_j+\epsilon_i \,\, \textup{or}\,\, r_{iT}\to r_{jT}+\epsilon_i.$ It readily follows from the above relation that $d_{ij}>0$, which represents the safety condition as in \eqref{eqn:coll_avoid}. The above condition ensures $d_{ij}>0, \,\,\forall \, t\geq 0$, while $\psi_{ij}<0$ and $\dot{\psi}_{ij}>0$ in Region IV of the $i$\textsuperscript{th} pursuer, as obtained in \eqref{eqn:colincondition} of \Cref{lem11}. This will result in $\psi_{ij}$ starting from a negative value to become zero. At this moment, $P_j$ will no longer be the nearest colliding neighbor for $P_i$. Further, if no other vehicle becomes the new nearest colliding neighbor, $P_i$ will undergo the behavior in \textbf{Mode I} to reach the desired proximity, settling behind $P_j$ and exhibiting the self-organizing behavior. Otherwise, $P_i$
    will undergo the behavior in \textbf{Mode II} to avoid collision with the new nearest colliding neighbor and settle behind it. Consequently, every pursuer starting away from the target will alternate between \textbf{Mode I} and \textbf{Mode II} behaviors, resulting in every pursuer to converge to circle of desired proximity without colliding with other pursuers and organising autonomously on enclosing shape. This concludes the proof.
    \end{proof}  
    The proposed guidance law \eqref{eqn:ctrl} is nonsingular even if the term $\sigma_i$ appears in its denominator, that is, the expression in \eqref{eqn:ctrl} never becomes unbounded if $\sigma_i\rightarrow 0\,\textup{or}\,\,\pm \pi$. In the steady state, $\dot{e}_i=-v\cos\sigma_i=0$ results in $\sigma_i \to \pm\pi/2$ as the equilibrium points.  Therefore, $\sigma_i=0\,\textup{or}\,\pi$ are not the equilibrium points for the pursuers utilizing the proposed guidance law. However, $\sigma_i= 0 \, \, \textup{or} \,\,\pi$ may occasionally occur in the transient phase during which $a_i$ will saturate momentarily to drive the system away from such configurations. However, during implementation, control inputs are constrained to remain within finite bounds since the control input values practically cannot be infinite. Therefore, whenever the demand for $a_i$ grows beyond the allowable limits, the control input will saturate and steer the pursuer away from $\sigma_i=  0 \, \, \textup{or} \,\,\pi$.     
\begin{remark}
    The control input designed in \Cref{thm1} is solely based on relative measurements, making it lucrative even in scenarios when global information is not available. The proposed guidance law for the $i$\textsuperscript{th} pursuer \eqref{eqn:ctrl} is independent of the other pursuers' motion, thereby exhibiting robustness to their movement.
\end{remark}
\begin{remark}
    Ensuring the expression of $K_i$ as after \eqref{eqn:ctrl} is greater than zero provides an additional condition on the selection of the control parameters, that is, $\Delta_i>r_s$. This simply implies that the region of the dominance of the repulsive term denoted by $\Delta_i$ is greater than the $i$\textsuperscript{th} pursuer's sensing radius, resulting in the repulsive field dominating within the sensing radius. {Additionally, the selected values of parameters $\Delta_i, \delta_i$ and $\lambda_i$ directly impacts the minimum allowable safe distance $\epsilon_i$ (as discussed in \Cref{repl_rem}) among the pursuers. A larger value of $\epsilon_i$ results in a lesser number of pursuers safely converging to the desired proximity and vice-versa at the steady state.} 
\end{remark}
\section{Simulation Results}
In this section, we demonstrate the efficacy of the proposed control law in autonomously organizing the pursuers to enclose a stationary target. In the following results, the target is stationary at the position $[x_T\; y_T]=[0,0]^\top$ m. All pursuers start at different distances from the target and move at a constant speed of $40$ m/s. The initial positions of pursuers are denoted by square markers in the subsequent trajectory plots. The desired proximity from the target is set as $r_d=100$ m for all the simulation results. The controller gains and other parameters are selected as follows: $K_i = 10$, $\lambda_i=0.9$, $\eta_i=70000$, $\Delta=100$, and $r_s=50$ m.
\begin{figure*}[h]
	\centering
	\begin{subfigure}[t]{0.49\linewidth}
    \centering    \includegraphics[width=\linewidth]{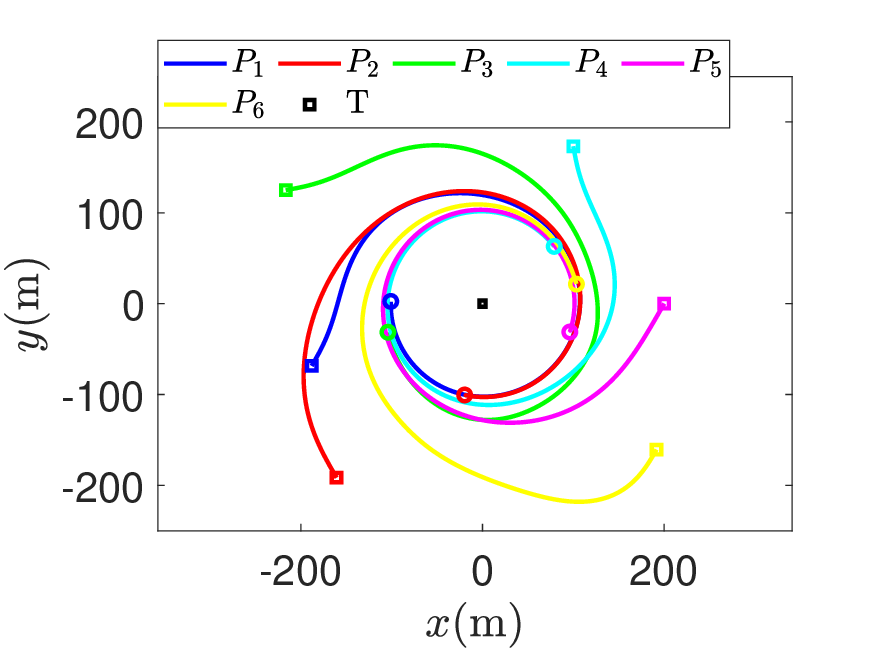}
    \caption{Trajectories.}
    \label{fig:traj_2p}
    \end{subfigure}
	\begin{subfigure}[t]{0.49\linewidth}
		\centering		\includegraphics[width=\linewidth]{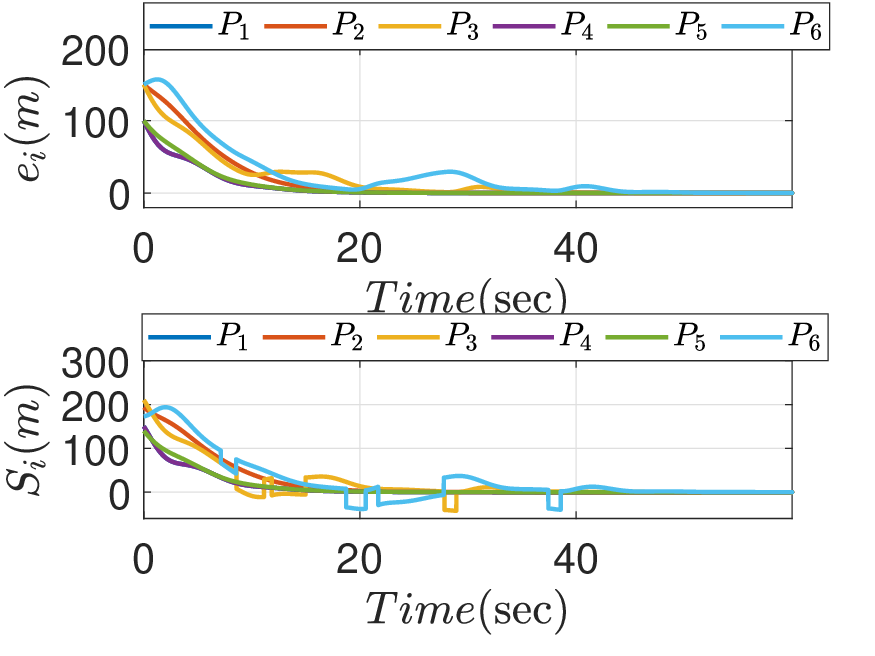}
		\caption{Error variables and sliding manifold.}
		\label{fig:e_2p}
	\end{subfigure}
	\begin{subfigure}[t]{0.49\linewidth}
		\centering
		\includegraphics[width=\linewidth]{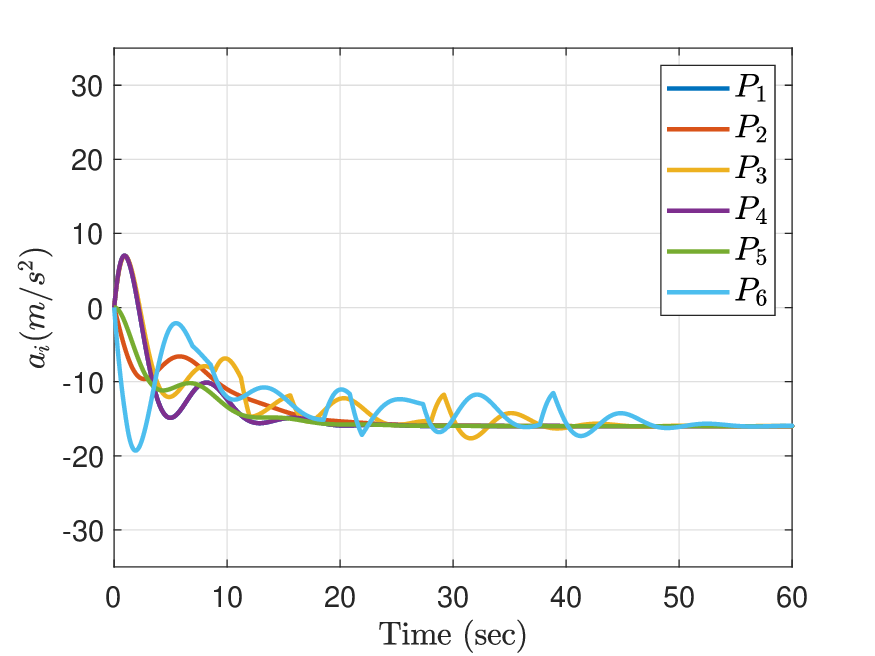}
		\caption{Lateral accelerations.}
		\label{fig:ctrl_2p}
	\end{subfigure}	
       \begin{subfigure}[t]{0.49\linewidth}
		\centering		\includegraphics[width=\linewidth]{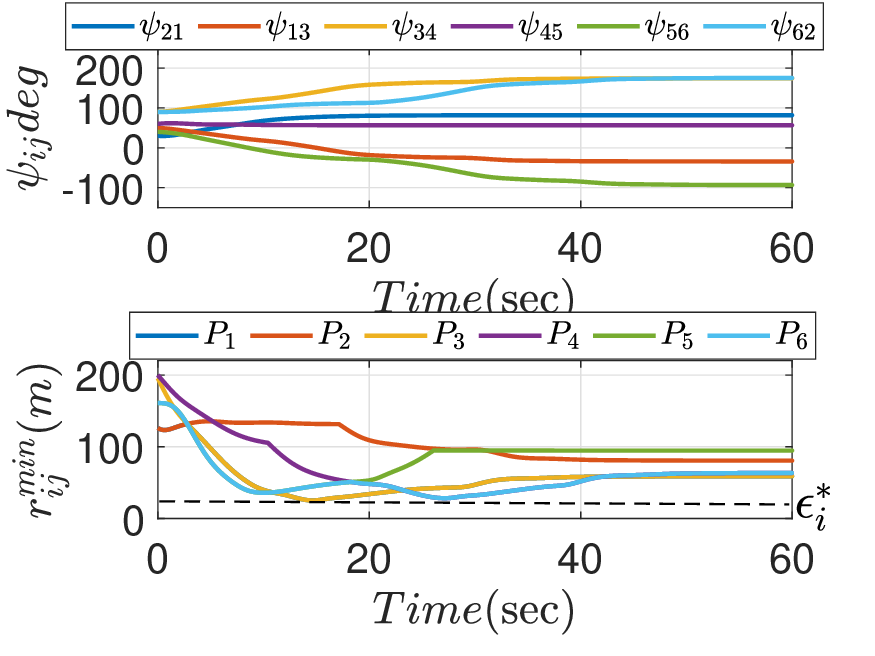}
		\caption{Inter-pursuer distances and angular spacings.}
		\label{fig:inter_2p}
	\end{subfigure}	
   	\caption{Collision-free self-organizing target enclosing with 6 pursuers.}
	\label{fig:2p}
\end{figure*}
\begin{figure*}[h]
	\centering
	\begin{subfigure}[t]{0.49\linewidth}
    \centering    \includegraphics[width=\linewidth]{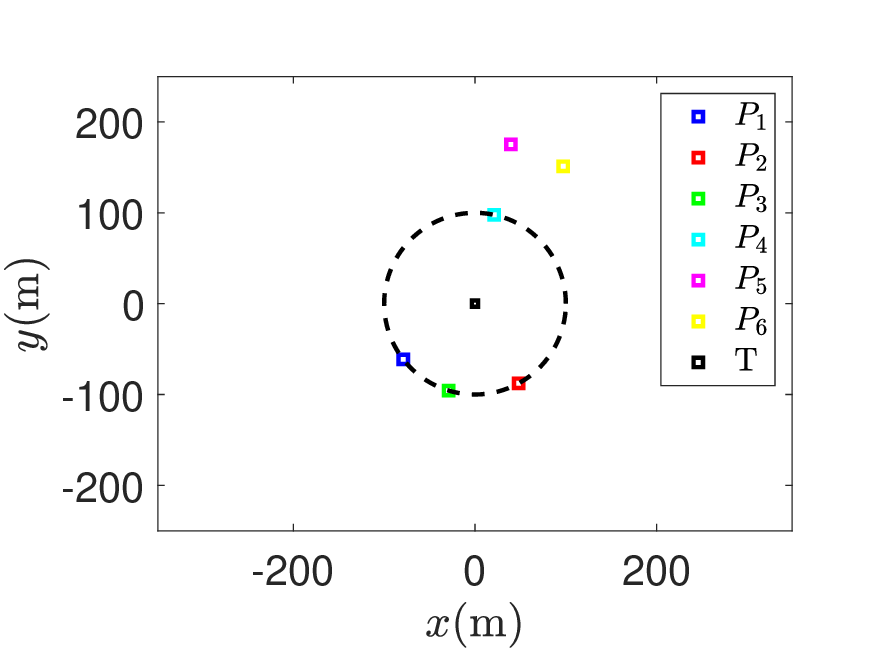}
    \caption{Trajectories ($t=0 \,sec$).}
    \label{fig:traj_2_1}
    \end{subfigure}
	\begin{subfigure}[t]{0.49\linewidth}
		\centering
		\includegraphics[width=\linewidth]{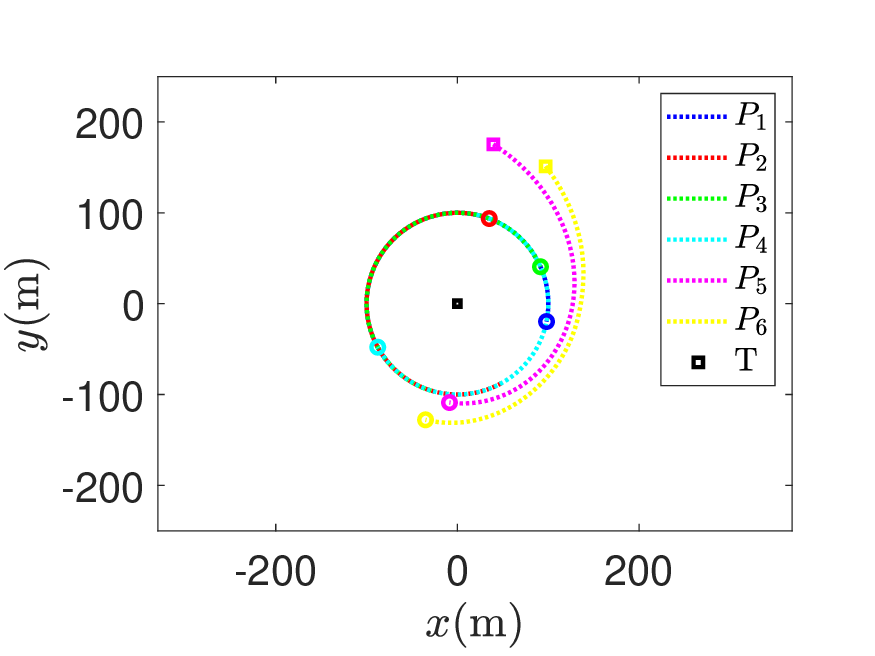}
		\caption{Trajectories ($t=15 \,sec$).}
		\label{fig:traj_2_2}
	\end{subfigure}
	\begin{subfigure}[t]{0.49\linewidth}
		\centering
		\includegraphics[width=\linewidth]{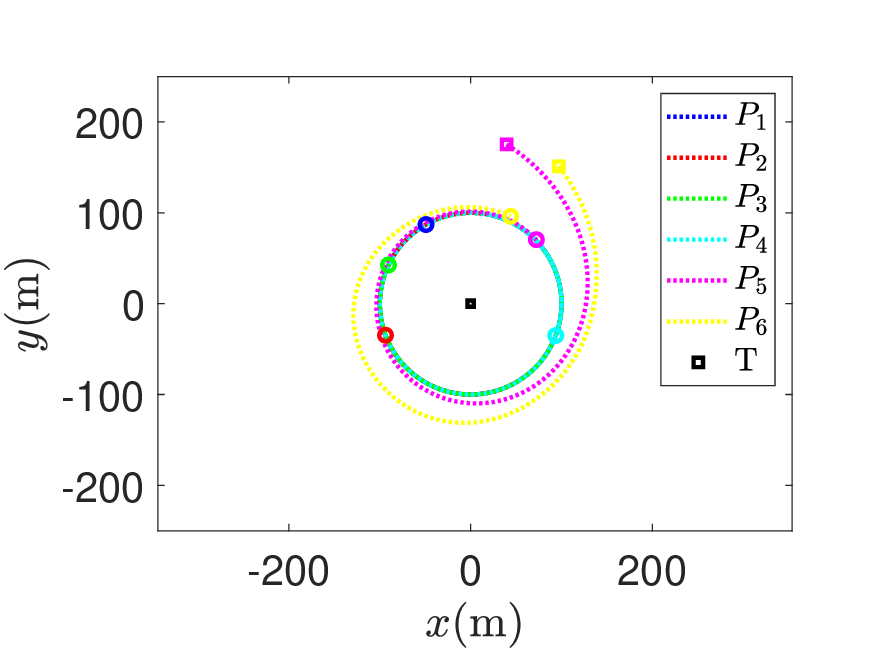}
		\caption{Trajectories ($t=35\, sec$).}
		\label{fig:traj_2_3}
	\end{subfigure}	
       \begin{subfigure}[t]{0.49\linewidth}
		\centering
		\includegraphics[width=\linewidth]{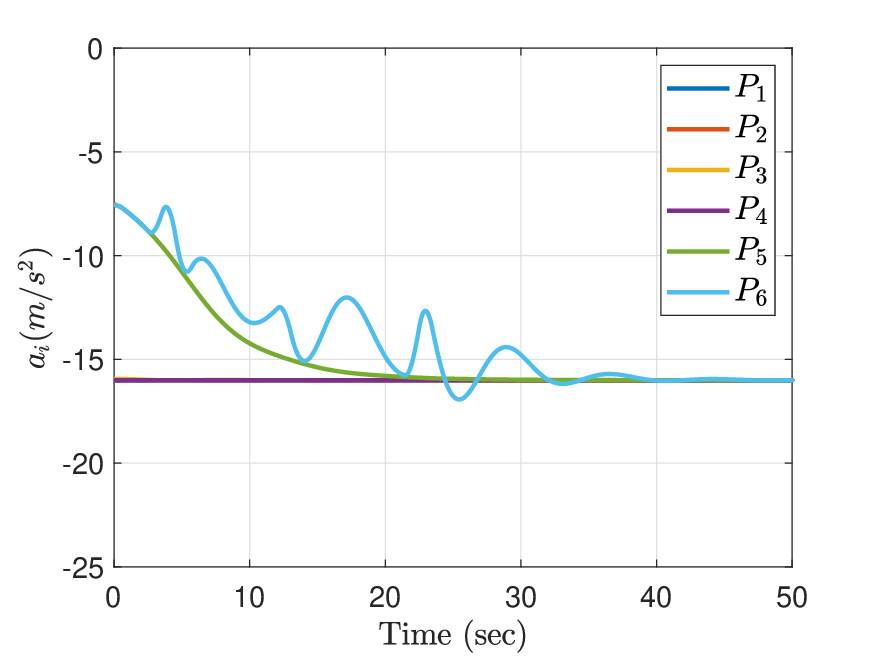}
		\caption{Lateral accelerations.}
		\label{fig:ctrl_2_1}
	\end{subfigure}	
   	\caption{Two pursuers joining the enclosing formation.}
	\label{fig:rel2}
\end{figure*}

The first set of results illustrates the scenario in which six pursuers start away from the target, as in \Cref{fig:2p}. The initial arrangement of the pursuers around the target in a clockwise direction is given as $P_2-P_1-P_3-P_4-P_5-P_6$, as shown by square markers in the figure. It is observed that the final arrangement of the pursuers on the desired proximity (as depicted by circular markers) is given as,  $P_2-P_3-P_1-P_4-P_6-P_5$. This indicates that only $P_3$ and $P_6$ encounter a nearest colliding pursuer and reorganize to avoid collision. Meanwhile, the $P_1, P_2, P_4,$ and $P_5$ move directly to the desired proximity without encountering the nearest colliding pursuer. The profiles of the error variables and sliding manifold are depicted \Cref{fig:e_2p}, where the sliding manifold converges to zero within a finite time. The profiles of $S_i$ for only $P_3$ and $P_6$ depict sudden deviations at different instances of time, which represent instances of activation/deactivation of the collision avoidance component in the control input. Note that whenever collision avoidance is activated, the range error diverges from zero momentarily to ensure safety. 

\Cref{fig:ctrl_2p} depicts the profiles of lateral accelerations of the pursuers converging to the same constant value at a steady state, as all the pursuers eventually converge to the same loiter circle of radius of desired proximity. The lateral acceleration profiles of $P_3$ and $P_6$ vary from the other pursuers due to the activation of the collision avoidance component. Finally, the profiles of inter-pursuer variables are shown in \Cref{fig:inter_2p}, where the profiles of $\psi_{ij}$ change sign crossing zero for only $P_3$ and $P_6$. This shows that these pursuers have autonomously reorganized from their initial configurations, where the profiles $r_{ij}^{\min}$ depict that the distance between the two closest pursuers never reaches zero. {Further, it can be observed that all the pursuers ensure $r_{ij}^{\min}>{\epsilon_i}^*\;\forall t\geq0$ (as depicted by the black dotted line on $r_{ij}$ profiles in \Cref{fig:inter_2p}) to maintain a safe distance with the pursuers converging to the desired proximity with only the attractive component the potential function in the control input, as presented in the controller design.} 

\Cref{fig:rel2} illustrates the second set of results where initially, four pursuers ($P_1$ to $P_4$) are already moving on the circle of desired proximity when two other pursuers ($P_5$ and $P_6$) start away from the desired proximity. From the \Cref{fig:traj_2_1,fig:traj_2_2,fig:traj_2_3}, it is evident that $P_5$ reaches the desired proximity without encountering a nearest colliding pursuer and autonomously organizing on the free space on the enclosing shape. While $P_5$ itself becomes the nearest colliding pursuer for $P_6$, resulting in $P_6$ taking a larger route to converge to the desired proximity behind $P_6$. This is verified by \Cref{fig:ctrl_2_1}, where only the profiles related to $P_6$ exhibit abrupt variation compared to other pursuers, indicating {the} activation of the repulsive component of the control input. Therefore, $P_5$ and $P_6$ are able to autonomously join the enclosing formation without colliding with other pursuers. Furthermore, we offer video demonstrations of our simulation results, accessible at \href{https://t.ly/x533O}{https://t.ly/x533O}. The videos titled ``results\_1.mp4" and ``results\_2.mp4" showcase the simulation outcomes corresponding to the scenarios presented in the paper. Additionally,``results\_3.mp4" and ``results\_4.mp4" illustrate the self-organizing target enclosing in large swarms with 10 and 20 pursuers, respectively.

\section{Conclusions}
In this paper, we proposed guidance laws for multiple pursuers to safely and autonomously organize at the desired proximity to encircle a stationary target, using only the relative information. In our approach, the agents are free to converge to any point on the enclosing shape as long as they maintain a safe distance between them, which may result in the pursuers not requiring a pre-determined formation structure and autonomously organizing on the enclosing shape. Our guidance strategy ensures that each pursuer selects at most one other pursuer based on certain collision conditions. We show that preventing collision with the selected other pursuer is sufficient to avoid collisions with all the other pursuers. To tailor the desired switching behavior required, we proposed a potential function such that the attractive component of the potential is activated all the time. Meanwhile, the repulsive component is activated whenever the pursuer encounters a nearest colliding pursuer. Further, we utilized sliding mode control to develop the reaching law, ensuring finite time convergence of the sliding manifold, followed by asymptotic convergence of the error to zero. The resulting guidance law enables pursuers to move towards the desired proximity along collision-free paths, offering increased maneuvering flexibility, requiring minimal information, and demonstrating robustness to the motion of other pursuers. Validation of the guidance laws with different numbers of pursuers attests to the merits of the proposed guidance algorithm. Incorporating a moving target and consideration of actual vehicle models, possibly in 3D space, could be interesting to pursue in the future. Additionally, the current work can be extended to include an in-depth investigation of the switching behavior and the effect of dynamic obstacles in the environment.

\bibliographystyle{ieeetr}
\bibliography{references}

 \end{document}